\documentclass[preprint]{aastex}
\usepackage{color, soul}
\usepackage{graphics}
\usepackage{subfigure}
\usepackage{rotating}
\usepackage{lscape}
\usepackage{cases}
\usepackage{gensymb}
\begin{document}

\shorttitle{Galactic Center {\it Chandra} Source Catalog}
\title{An Ultra-deep {\it Chandra} Catalog of X-ray Point Sources in the Galactic Center Star Cluster}
\author{Zhenlin Zhu\altaffilmark{1,2}, Zhiyuan Li\altaffilmark{1,2}, Mark R. Morris\altaffilmark{3}}
\affil{$^{1}$ School of Astronomy and Space Science, Nanjing University, Nanjing 210046, China}
\affil{$^{2}$ Key Laboratory of Modern Astronomy and Astrophysics (Nanjing University), Ministry of Education, Nanjing 210046, China}
\affil{$^{3}$ Department of Physics and Astronomy, University of California, Los Angeles, CA 90095, USA}
\email{zhuzl@smail.nju.edu.cn; lizy@nju.edu.cn}

\begin{abstract}
We present an updated catalog of X-ray point sources in the inner 500$\arcsec$ ($\sim$20 parsec) of the Galactic Center (GC), where the {\it nuclear star cluster} (NSC) stands, based on a total of $\sim$4.5 Ms of {\it Chandra} observations taken from September 1999 to April 2013.
This ultra-deep dataset offers unprecedented sensitivity for detecting X-ray sources in the GC, down to an intrinsic 2--10 keV luminosity of $1.0\times10^{31}{\rm~erg~s^{-1}}$.
A total of 3619 sources are detected in the 2--8 keV band, among which $\sim$3500 are probable GC sources and $\sim$1300 are new identifications.
The GC sources collectively account for $\sim$20\% of the total 2--8 keV flux from the inner 250$\arcsec$ region where detection sensitivity is the greatest.
Taking advantage of this unprecedented sample of faint X-ray sources that primarily traces the old stellar populations in the NSC, we revisit global source properties, including long-term variability, cumulative spectra, luminosity function and spatial distribution.
Based on the equivalent width and relative strength of the iron lines, we suggest that in addition to the arguably predominant population of magnetic cataclysmic variables (CVs), non-magnetic CVs contribute substantially to the detected sources, especially in the lower-luminosity group. 
On the other hand, the X-ray sources have a radial distribution closely following the stellar mass distribution in the NSC, but much flatter than that of the known X-ray transients, which are presumably low-mass X-ray binaries (LMXBs) caught in outburst. This, together with the very modest long-term variability of the detected sources, strongly suggests that quiescent LMXBs are a minor ($\lesssim$ a few percent) population. 
\end{abstract}

\keywords {Galaxy: center -- X-rays: stars -- X-rays: binaries}
\section{Introduction} {\label{sec:intro}}
Thanks to its proximity ($d \approx$ 8 kpc, 1$\arcsec$ corresponds to 0.039 pc; \citealp{Ghez2008,Gillessen2009}), the Galactic center (GC) offers a unique laboratory for studying the profound astrophysics in galactic nuclei, and thus is hotly pursued by multi-wavelength observations.
In particular, near-infrared (NIR) observations have resolved the most luminous members of the stellar populations in the GC (\citealp{Becklin1968}; \citealp{Eckart1993}; \citealp{Philipp1999}; \citealp{Schodel2007}; \citealp{Do2009}; \citealp{Wang2010}; \citealp{Dong2017}, among others), albeit necessarily through the heavily obscuring interstellar medium and a modest number of foreground interlopers. These resolved stars, mainly young massive stars and red giants, contain crucial information about the evolutionary history of the {\it nuclear bulge} and the interplay between the central super-massive black hole, also known as Sgr A*, and its immediate environment (see review by \citealp{Genzel2010}). 


X-ray observations of the GC, in particular those afforded by the {\it Chandra X-ray Observatory} (\citealp{Weisskopf2002}), provide an excellent complement to the NIR observations. Unlike the latter that primarily trace isolated, nuclear burning stars, the {\it Chandra} observations, with sensitivities down to $L_{\rm X} \lesssim 10^{32}{\rm~erg~s^{-1}}$, capture accretion-powered, close binary systems, in which a black hole (BH), a neutron star (NS), or a white dwarf (WD), accretes from a typically low-mass companion. 
Such close binaries represent the older stellar populations in the GC, and are also sensitive probes of the stellar dynamics in this dense environment, especially under the gravitational influence of Sgr A* (e.g., \citealp{Stephan2016}).
In addition, the X-ray observations have also revealed tens of Wolf-Rayet and O stars with strong winds (e.g, \citealp{Mauerhan2010}), and can place interesting constraints on the quantity of low-mass young stars (\citealp{Nayakshin2005}) as well as isolated BHs and NSs in the GC (e.g., \citealp{Deegan2007}).

The first {\it Chandra} survey of the GC, carried out by \citet{Wang2002}, covered a 2$\degree \times0.8\degree$ field along the Galactic plane in an array of 12-ks exposures.
This survey, albeit shallow, led to the recognition that the Fe\,XXV K$\alpha$ line in the GC predominantly arises from numerous faint ($L_{\rm X} \lesssim 10^{34}{\rm~erg~s^{-1}}$) point sources, rather than from a putative hot diffuse gas (\citealp{Koyama1989}).
Subsequently, \citet{Muno2003} reported on a total of 590 ks {\it Chandra} exposures toward the inner $\sim$20 parsecs, where the {\it nuclear star cluster} (NSC) stands (\citealp{Launhardt2002}), resolving $\sim$2000 sources that collectively account for $\sim$10\% of the total 2--8 keV emission from this region.
Later, Muno et al.~(2006, 2009) provided updated source catalogs in the 2$\degree \times0.8\degree$ field, based on increasingly deeper {\it Chandra} observations.

A leading candidate for the faint X-ray sources in the GC is magnetic cataclysmic variables (mCVs), primarily intermediate polars (IPs). This is supported by several lines of evidence. First, the surface number density of the sources with $L_{\rm X} \gtrsim 10^{32}{\rm~erg~s^{-1}}$ (50\% completeness) in the 2$\degree \times0.8 \degree$ field roughly follows the NIR starlight distribution (\citealp{Muno2009}, hereafter M09), and the X-ray source abundance (i.e., number of sources per unit stellar mass) is in rough agreement with the CV abundance in the Solar neighborhood (\citealp{Sazonov2006}).
Second, the faint sources exhibit in their spectra strong Fe\,XXVI Ly$\alpha$ and Fe\,XXV K$\alpha$ lines as well as an intrinsically hard continuum, which are typical of IPs, but unlikely from other known classes of X-ray sources of similar luminosities (\citealp{Muno2004}). Third, the extended 20--40 keV emission from the inner few parsecs of the GC, recently discovered by NuSTAR (\citealp{Perez2015}), is consistent with originating from hundreds to thousands of IPs having an average WD mass of $\sim$0.9 ${\rm M}_{\odot}$, which is in good agreement with the mean WD mass measured in field IPs (\citealp{Hailey2016}).   
Lastly, a handful of the GC X-ray sources shows flux modulations with periods ranging from a few hundred seconds to a few hours, which are again typical of mCVs (\citealp{Muno2004}; M09). 

In addition to the mCV candidates, which on average show little drastic flux variability, about a dozen transient sources have been discovered in the GC since the launch of {\it Chandra} and {\it XMM-Newton}, and more recently due to the {\it Swift} monitoring program of high cadence since 2006 (\citealp{Degenaar2012,Degenaar2015,Ponti2016}). 
The majority of these transients are proposed to be low-mass X-ray binaries (LMXBs), having interesting peak X-ray luminosities of $10^{34-36}{\rm~erg~s^{-1}}$, which, on the one hand, distinguish them from normal CVs, but on the other hand, are orders of magnitude lower than typical outburst luminosities of LMXBs found in the Galactic disk (\citealp{Wijnands2006}). 
For this reason, the suggestion has also been made that at least some of these transients are classical novae (\citealp{Mukai2008}).


The ever-increasing {\it Chandra} exposure toward the GC allows for detecting fainter sources and placing tighter constraints on the individual and global source properties. 
In this work, we present a new X-ray source catalog of the inner 20 pc, where the NSC resides, based on $\sim$4.5 Ms of {\it Chandra} observations, a dataset nearly doubling the temporal baseline achieved in the last major catalog of M09.
Section 2 describes the {\it Chandra} observations and our data preparation.
Section 3 is dedicated to the source detection procedure. 
Section 4 presents the main source catalog in detail, along with a brief comparison with the M09 catalog.
Section 5 addresses the nature of the GC X-ray sources, through statistical analyses of their global properties, including long-term variability, cumulative spectra, luminosity function and spatial distributions. 
Section 6 summarizes our results and discusses implications for the GC X-ray stellar populations. 
Throughout this work, we quote errors at 1\,$\sigma$ (68.3\%) confidence level, unless otherwise stated.

\section{Observations and Data Preparation}
\label{sec:data}
Sgr A* and its vicinity have been frequently visited by {\it Chandra}, primarily with its Advanced CCD Imaging Spectrometer (ACIS). 
We utilized a total of 87 {\it Chandra} observations, which all have their aim-point targeted within $\sim$1\arcmin\ from Sgr A*, ensuring an optimal point-spread function (PSF). 
A log of the {\it Chandra} observations used in this work is given in Table~\ref{tab:obs}.
Among these observations, 49 were taken with the ACIS-I array, spanning 13.5 years (September 1999 to April 2013).
We note that ObsID 1561 had two exposures separated by a few months and essentially independent of each other, hence we renamed them as ObsID 1561a and 1561b for clarity. 
The other 38 observations, taken in 2012, had the ACIS-S and High Energy Transmission Grating (HETG) in operation, chiefly to resolve the accretion flow onto Sgr A* (\citealp{Nowak2012}; \citealp{Wang2013}).
With the HETG inserted, of order half the incident X-rays are dispersed, while the remaining X-rays continue to the detector directly and form the ``zeroth-order" image.
It is noteworthy that since 2013 there have been quite a few non-grating ACIS-S observations pointed toward Sgr A*. However, these observations were taken in a subarray mode with a small field-of-view. Hence we did not include them in this work, at the price of losing track of a few of transients. 


We downloaded and uniformly reprocessed the archival data with CIAO v4.8 and the corresponding calibration files, following the standard procedure\footnote{http://cxc.harvard.edu/ciao}. 
Briefly, we used the CIAO tool {\sl acis\_process\_events} to reprocess the level 1 event files, applying both charge transfer inefficiency (CTI) and gain corrections, as well as filtering events by cosmic-ray afterglows.
All but one observation were taken in the FAINT mode (ObsID 242 in VFAINT).
The CIAO tool {\sl reproject\_aspect} was employed to calibrate the relative astrometry among the individual observations, by matching the centroids of commonly-detected point sources. ObsID 3392, which has the longest exposure among the ACIS-I observations, served as the reference frame.
The resultant accuracy in relative astrometry was typically better than $0\farcs1$.
After the level 2 event file was created for each ObsID, we constructed a merged event list, reprojecting all events to a common tangential point, i.e., the position of Sgr A* ([$RA$, $DEC$]=[17:45:40.038, -29:00:28.07]).
We have also examined the light curve of each ObsID and found that 
only mild particle flares were present in $\lesssim$1\% of the total time intervals, which have a negligible contribution ($\lesssim 5\times10^{-4}$) to the total background. Hence we decided to preserve all the science exposures for source detection and characterization, maximizing the useful signals.
The total exposure amounts to 4.44 Ms (1.50 Ms from ACIS-I and 2.94 Ms from ACIS-S/HETG).

For each observation, we generated counts maps and exposure maps in four energy bands: 0.5--2, 2--3.3, 3.3--4.7 and 4.7--8 keV.
We included only data from the I0, I1, I2 and I3 CCDs for the ACIS-I observations, and only data from the S2 and S3 CCDs for the ACIS-S observations, to ensure optimal sensitivity for source detection.
The exposure maps were weighted by an absorbed bremsstrahlung spectrum, with a plasma temperature of 10 keV and an absorption column density of $N_{\rm H}$ = 1.0$\times10^{23} {\rm~cm^{-2}}$, which is empirically appropriate for most GC X-ray sources (Section \ref{sec:hr}).
PSF maps of a given enclosed count fraction (ECF) were also produced using the same spectral weighting. 
The individual exposure maps were then reprojected to form a combined exposure map of a given band.
The individual PSF maps of a given band were similarly combined, weighted by the exposure.

Figure \ref{fig:tricolor} displays a false-color image of the inner 500$\arcsec$ of the GC, with 2--3.3, 3.3--4.7 and 4.7--8 keV intensities shown in red, green and blue, respectively.
We note that the central 250$\arcsec$ is covered by all observations, while regions between 250$\arcsec$ and 500$\arcsec$ are only fully covered by the ACIS-I observations.
The use of HETG in the ACIS-S observations necessarily disperses photons into the field-of-view. 
This is most clearly seen in the greenish regions in Figure~\ref{fig:tricolor}, which can be attributed to first-order dispersed 3.3--4.7 keV photons from the X-ray-bright Sgr A complex (\citealp{Baganoff2003}).  
A complete removal of such dispersed photons is impractical, but fortunately they would only contaminate the diffuse background on scales much larger than the PSF. 
Our rough estimate indicated that this contamination elevates the background level in the greenish regions by a factor of 1.4, where the signal-to-noise ratio (S/N) for point sources is more than compensated by including the $\sim$3 Ms ACIS-S exposure.  
A comparison between the ACIS-S and ACIS-I photometry (Section \ref{sec:pho}) also suggests that contamination from the dispersed photons has little effect in the source photometry, as long as the local background is properly accounted for.

To gain insight on the differences and similarities between the X-ray source populations in the nuclear bulge and the Galactic bulge, we also utilized 13 ACIS-I observations of the so-called {\it Limiting Window} (LW; Table \ref{tab:obs}),
a field sampling the Galactic bulge at a projected distance of $\sim$80$\arcmin$ from Sgr A*, against a relatively low line-of-sight column density of $N_{\rm H} \approx 7\times10^{21}{\rm~cm^{-2}}$ (\citealp{Revnivtsev2011}).
These observations, in a total exposure of 980 ks, resolved $\sim$80\% of the 6.5--7.1 keV flux into discrete sources, providing compelling evidence that the so-called Galactic Ridge X-ray Emission has a predominantly stellar origin (\citealp{Revnivtsev2009}).
We downloaded the data and produced combined images of the LW following the same procedure as for the GC.  

\section{Source Detection}
\label{sec:detection}
Since the first systematic effort of \citet{Muno2003}, it has been recognized that identifying and characterizing X-ray point sources in the GC is a challenging task, due to the high source density, strong and non-uniform diffuse background, and severe foreground extinction. We carried out source detection and pre-screening in the following steps, which share similar merits with M09. 

(i) We employed the {\it wavdetect} algorithm on the combined counts images in the 0.5--2 keV (soft) and 2--8 keV (hard) bands, adopting a false-positive probability threshold of 10$^{-6}$ and a ``2-sequence" of wavelet scales (i.e., 1, 2, 4 and 8 ACIS pixels).
A combined 50\%-ECF PSF map and combined exposure map were supplied to the detection process. 
A raw source list resulted for each band. 

Due to the high column density along the line-of-sight, essentially all photons with energies below $\sim$2 keV from the GC would be absorbed. 
Therefore, nearly all sources detected in the soft-band should lie in the foreground, i.e., in front of the nuclear bulge, except for the brightest ones (Section~\ref{sec:hr}). 
We present in Table \ref{tab:soft} a catalog of the 421 soft-band sources, but perform no in-depth analysis for them in this work.  
Source photometry, as described in Section \ref{sec:pho}, has been performed to derive the observed 0.5-2 keV photon flux ($S_{0.5-2}$).

(ii) The inner 20 pc of the GC is rich in extended X-ray features (Figure~\ref{fig:tricolor}). Some of these features, however, could have been picked up by {\it wavdetect}, due to the nature of the detection algorithm that searches for ``local peaks". 
We tried several semi-automated ways to flag extended sources in the hard-band raw list, but eventually came to the conclusion that visual examination is probably the most robust way, as was also adopted by M09.
Two of us (Z.Z. and Z.L.) independently carried out the exercise, and 199 extended sources visually identified by both of us were thus removed.
These fall into two main classes: diffuse clumps in the Sgr A East supernova remnant (\citealp{Maeda2002}), and filaments (e.g., \citealp{Muno2008}). The two well-known pulsar wind nebula candidates, the ``Cannonball" (\citealp{Park2005}) and G359.95-0.04 (\citealp{Wang2006}), as well as Sgr A* (\citealp{Wang2013}), were also regarded as extended sources and removed. For reference, we tabulate the positions of the extended sources in the Appendix, where we also provide an illustration of some examples.

(iii) To improve the reliability of the hard-band sources, we calculated the binomial no-source probability, $P_{B}$, defined as the probability of still observing the same number of counts or more assuming that the observed source is due to background fluctuations \citep{Weisskopf2007}: 
\begin{equation}
P_{B}(X \ge C_S) = \sum_{X=C_S}^{N} \frac{N!}{X!(N-X)!}p^{X}(1-p)^{N-X},
\end{equation}
where $C_S$ is the number of counts in the source-extraction region, 
$C_B$ is the number of counts in the background-extraction region, and $N = C_S + C_B$.
The parameter $p$ is defined as $p = A_{S}/(A_{S}+A_{B})$, A$_{S}$ and A$_{B}$ being the area of the source-extraction and background-extraction regions, respectively. 
The determination of $C_S$, $C_B$, $A_S$ and $A_B$ followed the photometry procedure (Section~\ref{sec:pho}). 
The distribution of $P_{B}$ versus the observed 2-8 keV photon flux is shown in Figure~\ref{fig:pb}.
We adopted a conservative threshold of $P_{B} > 0.1$ to filter probable background fluctuations.
In this way, 377 sources were removed from the hard-band raw list. For reference, we provide the positions of these spurious sources in the Appendix, along with an illustration of some examples. 
As evident in Figure~\ref{fig:pb}, the majority of these filtered sources are very faint and well below the 50\% completeness limit (Section \ref{sec:complete}). Any subtle ambiguity in this filtering would have little effect in our statistical analysis of the global source properties (Section 5).   

(iv) Lastly, we excluded from further analysis sources located at a projected distance $R >$ 500\arcsec\ from Sgr A*. Any such sources suffer from a large PSF and are only present in a small fraction of the observations.   

Using a similar procedure, we detected 846 sources in the 2--8 keV image of the LW. 
Since our main purpose is to contrast the global source properties between the GC and the Galactic bulge (Sections~\ref{sec:spec} and \ref{sec:LF}), we do not attempt to provide a full source catalog of the LW here.
X-ray sources in the LW, based on the same dataset used here, have been studied in detail by \citet{Revnivtsev2009}, \citet{Hong2012} and \citet{Morihana2013}.
\section{Main Catalog}
The final catalog of 2--8 keV point sources in the $R \leq 500\arcsec$ region is presented in Table \ref{tab:main}, which contains 3619 sources.
Figure \ref{fig:2d} displays the spatial distribution of these sources, along with those detected in the soft-band. 
We describe the catalog content in Section~\ref{sec:detail}, followed by our detailed procedures to characterize the individual sources.  

\subsection{Catalog Content}
\label{sec:detail}
Table \ref{tab:main} has the following columns:

\noindent Column (1): Source sequence number, in order of increasing Right Ascension.\\
Columns (2)--(3): Right Ascension and Declination (J2000) of the source centroid (Section \ref{sec:pu}).\\
Column (4): Positional uncertainty, in units of arcseconds (Section \ref{sec:pu}).\\
Column (5): The total number of 2--8 keV counts in the source-extraction region.\\
Column (6): The number of 2--8 keV background counts, normalized by the background-extraction area. \\
Column (7): 2--8 keV net counts and 1\,$\sigma$ uncertainties. \\
Column (8): Observed 0.5--2 keV photon flux, in units of $10^{-7}{\rm~photon~cm^{-2}~s^{-1}}$. In cases of non-detection in this band, a 1\,$\sigma$ upper limit is provided.\\
Column (9): Observed 2--8 keV photon flux, in units of $10^{-7}{\rm~photon~cm^{-2}~s^{-1}}$. \\
Column (10): Absorption-corrected 2--10 keV energy flux, in units of $10^{-15}{\rm~erg~cm^{-2}~s^{-1}}$. Columns (5)--(10) are derived in Section \ref{sec:pho}.\\
Columns (11)--(13): The soft, medium and hard colors and 1\,$\sigma$ uncertainties (Section \ref{sec:hr}). \\
Column (14): Notes on specific sources (Sections \ref{sec:m09} and \ref{sec:multi}). `m' for sources with counterparts in M09, `ir' for sources with NIR counterparts in \citet{Mauerhan2010}, `x' for X-ray transients tabulated in \citet{Degenaar2015}, and `f' for probable foreground sources.

\subsection{Source Position and Uncertainty}
\label{sec:pu}
We refined the {\it wavdetect} source centroids, using a maximum likelihood method \citep{Boese2001} that iterates over the recorded positions of the individual counts within the 75\% enclosed counts radius (ECR).   
In the rare case that the offset between the refined and original centroids was larger than the PSF, a neighboring source might have caused confusion, and we retained the original centroid of {\it wavdetect}.
We further found that application of this method made no significant improvements for sources with off-axis angles larger than about 250$\arcsec$, for which we also retained the original centroid. 

The positional uncertainty output by {\it wavdetect} could be underestimated, especially for faint sources. 
We followed the simulation results of \citet{Kim2007} to estimate the positional uncertainty ($PU$), at 68\% confidence level, 
as a function of net counts ($C_{\rm N}$) and off-axis angle ($OAA$), 
 \begin{numcases}{{\rm log} PU =} 
      0.1137\,OAA-0.4600\,{\rm log}C_{\rm N} -0.2398,  & log$C_{\rm N} \le $ 2.1227 \\0.1031\,OAA -0.1945\,{\rm log}C_{\rm N} -0.8034,  & 2.1227 $<$ log$C_{\rm N} \le$ 3.3000
\end{numcases}
where $PU$ is in units of arcseconds and $OAA$ in arcminutes. 
For a given source in the combined image, we determined its $OAA$ as the exposure-weighted mean of the source's $OAA$s in the individual observations.
For the few bright sources having more than 10$^{3.3}$ counts, i.e., beyond the applicable range of Equation 3, we still calculated their $PU$ following this formula.

\subsection{Photometry}
\label{sec:pho}
It is recognized that {\it wavdetect} tends to underestimate the net flux of faint sources \citep{Kim2007}, 
hence we performed aperture photometry for the detected sources.  
Although the {\it Chandra} PSF becomes substantially elliptical at large off-axis angles, we adopted a circular aperture, which is suitable for the case of combining multiple observations and empirically yields consistent results with elliptical aperture (e.g., \citealp{Wang2016}).
Source crowding in the GC necessitates caution; 
to extract the source counts in each observation, we adopted a 90\%-ECR aperture as default, but applied a 75\%-ECR aperture for 742 sources and 50\%-ECR for another 384 sources to avoid overlapping of the source-extraction regions. 
The background-extraction region of a given source was chosen to be a concentric annulus with inner-to-outer radii of 2--4 times the 90\%-ECR, excluding any pixels falling within 2 times the 90\%-ECR of any neighboring sources. 
Total and background counts were extracted and summed over all ObsIDs in which the source is covered, in the 0.5--2, 2--3.3, 3.3--4.7 and 4.7--8 keV bands. 
The net photon flux was then derived by correcting for the effective exposure and aperture. 
A by-product of this procedure is the net counts in individual observations, which contain information on source variability (Section~\ref{sec:var}).
The 2--8 keV photon flux ($S_{2-8}$) was derived by summing the three subbands, 
with errors estimated using a Bayesian algorithm \citep{Park2006}, BEHR\footnote{http://hea-www.harvard.edu/AstroStat/BEHR/}. 
Based on the best-fit model of the cumulative source spectrum (Section~\ref{sec:spec}), we adopted a photon flux-to-energy flux conversion factor of $2.38\times10^{-8}{\rm~erg~photon^{-1}}$, to calculate the 2--10 keV unabsorbed energy flux. 
For completeness, we have also performed photometry for sources detected in the 0.5--2 keV band (Table \ref{tab:soft}).

Figure \ref{fig:is} compares the 2--8 keV photon fluxes measured from the ACIS-I observations with those from the ACIS-S observations, for sources located at $R < 250\arcsec$.  
No significant bias can be seen between the two sets of photometry,
indicating a satisfactory calibration between the ACIS-I and ACIS-S data.

\subsection{Hardness Ratio}
\label{sec:hr}
We defined three hardness ratios, i.e., a soft color (HR0) between 2--3.3 and 0.5--2 keV, a medium color (HR1) between 3.3--4.7 and 2--3.3 keV, and a hard color (HR2) between 4.7--8 and 3.3--4.7 keV, 
following HR $= (C_{H}-C_{S})/(C_{H}+C_{S})$, where $C_{H}$ and $C_{S}$ are the net counts in the higher and lower energy bands, respectively. 
The three colors and their 1\,$\sigma$ uncertainties, derived with BEHR, 
are listed in Table \ref{tab:main}. 

The distribution of the three colors is shown in Figure \ref{fig:hr}, along with the soft color calculated for the 421 sources detected in the 0.5--2 keV band (red histogram in Figure~\ref{fig:hr}b; Section~\ref{sec:detection}).  
Eighty-eight of the soft-band sources were also detected in the hard-band (Figure~\ref{fig:hr}a). The brightest 9 sources among them have a 2--8 keV photon flux greater than $5\times10^{-6}{\rm~photon~cm^{-2}~s^{-1}}$ and HR0 $> 0.55$, which are most likely located in the GC.  
The remaining 79 sources are classified as foreground and denoted with `f' in Column (14) of Table~\ref{tab:main}.
The foreground sources, now amounting to 412 and marked as open circles in Figure~\ref{fig:2d}, have a more-or-less flat distribution across the field-of-view, as expected.
In Figure \ref{fig:hr}c,d, the medium and hard colors are contrasted with predicted hardness ratios of certain absorbed bremsstrahlung models, with varied absorption columns and temperatures. 
The median value of HR1 is quite compatible with the fiducial source spectrum, i.e., $N_{\rm H}=10\times 10^{22}$ cm$^{-2}$ and $kT = 10{\rm~keV}$. The somewhat higher observed values of HR2 can be explained by the presence of Fe lines between 6--7 keV (Section~\ref{sec:spec}), which are not accounted for by the simple bremsstrahlung model.

\subsection{Sensitivity and Completeness}
\label{sec:complete}
We used the CIAO tool {\it lim\_sens} to determine the detection sensitivity (\citealp{Kashyap2010}) across the field-of-view, adopting S/N = 4.89 to accommodate the false-positive probability of $10^{-6}$ in {\it wavdetect}. 
The input exposure, PSF and background maps were identical to those used in Section~\ref{sec:detection}. 
The median sensitivity as a function of projected radius from Sgr A* is shown as a black curve in Figure~\ref{fig:cm}a.
The ultra-deep data have allowed us to reach a sensitivity of 0.9$\times10^{-7}{\rm~photon~cm^{-2}~s^{-1}}$ in the inner 100$\arcsec$, gradually increasing to a few times $10^{-7}{\rm~photon~cm^{-2}~s^{-1}}$ at large radii.
To facilitate a direct comparison with the sensitivity that would have been achieved by M09 if an identical detection procedure had been adopted, we selected the same set of ACIS-I observations used by M09 (33 observations, with a total exposure of 1.0 Ms) to produce a sensitivity map. As shown in Figure \ref{fig:cm}a, our catalog achieves a factor of $\sim$2 better sensitivity in the inner 100\arcsec, and still a factor of $\sim$1.5 better around 250\arcsec, compared to M09 (red curve).

To assess the detection completeness in our main source catalog, we relied on MARX\footnote{http://space.mit.edu/cxc/marx-5.0/} simulations of artificial sources that would be recovered by the same {\it wavdetect} procedure. 
We divided the simulations into two parts: the $R < 250\arcsec$ region and $250\arcsec< R <500\arcsec$ region. 
For the inner region covered by all 87 observations, the ACIS-S exposure of 2.94 Ms can be effectively converted into an ACIS-I exposure of 1.21 Ms, according to the fiducial source spectrum. Therefore, we set the nominal simulated exposure as 2.71 Ms for the inner region, and 1.50 Ms for the outer region. 
We note that in principle the MARX simulations should be performed for the individual observations to reflect the varying roll angle. We have simulated the combined exposure at a fixed, exposure-weighted roll angle to save computational effort, which should be a reasonable choice. 

We assigned one of the following values for the input 2--8 keV photon flux: 1, 4, 7, 10 and 20$\times10^{-7}$ photon~cm$^{-2}$~s$^{-1}$. 
The input source spectrum was again the fiducial absorbed bremsstrahlung model. 
In each simulation run, we randomly placed 200 artificial sources at a given photon flux.
For both the inner and outer regions, we used 100 realizations for a given source position. Our entire simulations thus involved a total of $4\times10^{4}$ artificial sources, about an order of magnitude greater than the number of the real sources. 
To mimic the diffuse background, we adopted the background image output from {\it wavdetect}, but with kernel scales of ``1 2'' only, to preserve more detailed structures of the background.
Then, we ran {\it wavdetect} on the simulated images (i.e., artificial sources plus background), using exactly the same false-positive probability threshold, kernels, combined exposure maps and PSF maps as in Section \ref{sec:detection}. 

The fraction of artificial sources recovered by {\it wavdetect} and similarly filtered by the binomial no-source probability (Equation 1) was then counted, which reflects the degree of completeness at a given 2--8 keV photon flux and is shown in Figure \ref{fig:cm}b.  
The 90\% completeness for regions of $R < 100\arcsec$, $R < 250\arcsec$ and $250\arcsec < R < 500\arcsec$ is found at a flux of $5.5 \times 10^{-7}$, $3.5\times 10^{-7}$, and $8 \times 10^{-7}$ photon~cm$^{-2}$~s$^{-1}$, respectively.
We note that the X-ray-bright Sgr A complex is the cause of the somewhat poorer completeness at $R < 100\arcsec$ than in regions immediately outside (Figure~\ref{fig:2d}).

\subsection{Comparison with the M09 Catalog}
\label{sec:m09}
Our main catalog contains 3540 X-ray sources that are most likely located in the nuclear bulge. 
For comparison, the M09 catalog contains 3306 sources (after excluding sources classified as foreground) within the same $R < 500\arcsec$ region. 
At face value, the number of newly identified sources in our catalog, as indicated by the difference between these two numbers, seems to be incompatible with the much higher sensitivity achieved in this work (Figure \ref{fig:cm}a).
We take a closer look into this issue by cross-correlating the two catalogs, following the method of \citet{Hong2009}. 
For a given source in the M09 catalog and the closest source found in our catalog, their relative distance, $d_{r}$, is defined as the ratio of the angular offset between the two sources to the quadratic sum of the positional uncertainties.
The distribution of $d_{r}$ for all the M09 sources, shown in Figure \ref{fig:cc}a, naturally exhibits a bimodal shape. The first peak is formed by true counterparts between the two catalogs, while the second peak is dictated by 
the source surface density in the field-of-view.  
We adopt the local minimum between the two peaks, i.e., $d_{r} \le 3.0$, as a natural cut for genuine counterparts between the two catalogs.
This results in 2213 matches, which is denoted with `m' in Column (14) of Table \ref{tab:main}.
The remaining $\sim$1300 sources in our catalog can be considered new identifications, which are expected given the much higher sensitivity and the relatively steep luminosity function of the GC X-ray sources (Section \ref{sec:LF}).



Meanwhile, the above exercise indicates that $\sim$1100 M09 sources within $R = 500\arcsec$ have no apparent counterpart in our catalog.
This discrepancy can be attributed to several factors:
i) A few tens of extended sources identified by us were regarded as point sources in M09. This is particularly the case for the clumps in Sgr A East; 
ii) Fluctuations in the diffuse background, which might have been misidentified as true sources by M09, would be largely suppressed in our combined image, especially within the inner 250$\arcsec$;
iii) In addition to {\it wavdetect}, M09 had also employed the {\it wvdecomp} algorithm \citep{Vikhlinin1998} to detect faint sources, effectively using a lower S/N threshold. Such an approach, however, has been called into question in the case of the LW \citep{Hong2009}. 
Indeed, the majority of those M09 sources without a counterpart in our catalog are faint sources, with fluxes below $\sim$$3\times 10^{-7}$ photon~cm$^{-2}$~s$^{-1}$. A visual examination in our combined image suggests that at the positions of these sources there are no obvious ``peaks" over the local background (see examples in the Appendix); 
iv) Intrinsic downward variability might also make some of the M09 sources undetected even in our deeper data, although this is unlikely to be the main cause.
We note that among the fainter sources in our catalog, up to 33 might be background fluctuations, according to the compound binomial no-source probability (Figure~\ref{fig:pb}). 
However, we emphasize that our statistical analyses largely avoid using the faintest sources (Section~\ref{sec:global}), and thus our results are not affected by the residual background fluctuations. 

For the 2213 common sources, we compare in Figure \ref{fig:cc}b their 2--8 keV photon fluxes given by the two catalogs. 
No substantial bias can be seen in the photometry, despite subtle difference in the analysis procedure between the two works, which is difficult to fully quantify.

\subsection{Multi-wavelength Counterparts}
\label{sec:multi}
For future reference, we provide cross-correlations with notable source catalogs, based solely on positional coincidence.
In particular, \citet{Mauerhan2010} identified 31 NIR sources with significant X-ray emission, which are probably massive stellar binaries with strong colliding winds.   
Among them, nine are detected in the inner 500$\arcsec$ and denoted with `ir' in Column (14) of Table~\ref{tab:main}. 
We also use `s' to denote 10 X-ray transients tabulated in \citet{Degenaar2015}, which all have a counterpart in our catalog. 
We note that the Galactic center magnetar, SGR\,J1475-29, was first discovered by {\it Swift} (\citealp{Degenaar2013}) as a transient ten days after the last {\it Chandra} observation (ObsID 14942) used here, and hence is not included in our catalog.  
The same is true for two transients captured in 2016 (\citealp{Degenaar2016}; \citealp{Ponti2016}).

\section{Global Source Properties} {\label{sec:global}}
In the following we investigate global source properties, including variability, flux distribution, spatial distribution and cumulative spectrum, which hold promise for revealing the still elusive nature of the GC X-ray sources.
While these statistical properties have been studied in various previous works, most notably Muno et al.~(2003, 2006, 2009), \citet{Revnivtsev2007} and \citet{Hong2009}, the unprecedented temporal baseline and sensitivity achieved in the present work should provide much stronger constraints for the sources in the NSC. 
 
\subsection{Variability}
\label{sec:var}

To quantify the flux variability among the 87 observations with temporal separations from days up to 13.5 years,
we define for each source in the main catalog (excluding the foreground sources) a variability index, VI =$ S_{max}/S_{min}$, where $S_{max}$ and $S_{min}$ are the maximum and minimum 2--8 keV photon fluxes measured among individual observations that cover the source (Section~\ref{sec:pho}).
This requires that the source was significantly detected in at least two observations.
Two faint sources do not meet this criterion and are excluded. 

The distribution of VI as a function of $S_{max}$ is shown in Figure \ref{fig:var}.
Thirteen sources exhibit strong variability, i.e., VI $> 100$. 
Among these highly variable sources, 8 are known transients: AX\,J1745.6-2901, CXOGC\,J174535.3-290124, CXOGC\,J174540.0-290005, CXOGC\,J174538.0-290022, {\it Swift}\,J174535.5-285921, CXOGC\,J174540.0-290031, CXOGC\,J174541.0-290014, XMMU\,J174554.4-285456 (\citealp{Degenaar2015}).
The other five sources seem to have drawn little attention before, probably because their maximum fluxes were not as high. 
The brightest among them, CXOGC\,J174537.6-290035 (source No.1857 in Table~\ref{tab:main}), was caught in outburst in ObsID 6363 (see also M09). We find that its spectrum during outburst can be fitted by an absorbed power-law model, with a photon-index of $1.6^{+1.5}_{-1.2}$ and $N_{\rm H} = 1.4^{+0.8}_{-0.7}\times10^{23}{\rm~cm^{-2}}$, giving an unabsorbed 2--10 keV luminosity of $\sim$$8\times10^{33}{\rm~erg~s^{-1}}$.
These values are not unlike those reported in the literature for the other (albeit somewhat brighter) transients, suggesting that CXOGC\,J174537.6-290035 could be an LMXB at the GC. 

The histogram of VI peaks at a value of $\sim$7 (Figure \ref{fig:var}). At a glance, this seems to indicate substantial variability in most sources.  
However, pure Poisson fluctuations could mimic variability in relatively faint sources.
To test this possibility, we simulate the distribution of VI for 6400 constant sources with Poisson fluctuations as would be realized in the 87 ACIS exposures, ignoring background contribution.
The simulated sources assume one of three intrinsic photon fluxes, $5\times 10^{-7}$, $1\times10^{-6}$ or $5\times10^{-6}$ photon~cm$^{-2}$~s$^{-1}$, with a relative number proportional to (photon flux)$^{-1.5}$, to reflect a power-law luminosity function (Section \ref{sec:LF}). 
The resulting VI distribution, shown as the black histogram in Figure~\ref{fig:var}, peaks at a value similar to that of the real sources. 
This rather simplified exercise suggests that the apparent variability in most of the faint sources can be explained by statistical fluctuations. 
On the other hand, sources with VI $\gtrsim 15$, which account for 14.6\% of all sources, probably exhibit some intrinsic variability. 
We defer a detailed study of the source variability, both short-term and long-term, to a future work.

\subsection{Cumulative Spectra}
\label{sec:spec}
The mild variability in most faint X-ray sources facilitates the quantification of their mean spectral properties.
It has been shown that the cumulative source spectrum exhibits an intrinsically hard continuum and multiple emission lines, especially the neutral Fe K$\alpha$ (6.4 keV), Fe XXV K$\alpha$ (6.7 keV) and Fe XXVI Ly$\alpha$ (7.0 keV) (\citealp{Muno2004}).
The flux ratio and equivalent width (EW) of these lines are useful diagnostics of the responsible source populations (e.g., \citealp{Xu2016}).

It is known that the effect of CTI degrades the spectral resolution of ACIS, in the sense that signals recorded at smaller detector rows have a poorer resolution\footnote{http://cxc.cfa.harvard.edu/proposer/POG/html/index.html}.
Therefore, approximately speaking, regions close to Sgr A* suffer from a poorer spectral resolution as recorded in the ACIS-I observations than in the ACIS-S observations.
Hence we define the following four sets of cumulative spectra, to ensure an optimal resolution for the Fe lines.
 
\noindent Set I1: 739 sources with $R > 250\arcsec$ and $S_{\rm 2-8} > 3.5\times10^{-7}{\rm~photon~cm^{-2}~s^{-1}}$;\\
Set I2: 348 sources with $R > 250\arcsec$ and $S_{\rm 2-8} < 3.5\times10^{-7}{\rm~photon~cm^{-2}~s^{-1}}$;\\
Set S1: 594 sources with $R < 250\arcsec$ and $S_{\rm 2-8} > 3.5\times10^{-7}{\rm~photon~cm^{-2}~s^{-1}}$;\\
Set S2: 1788 sources with $R < 250\arcsec$ and $S_{\rm 2-8} < 3.5\times10^{-7}{\rm~photon~cm^{-2}~s^{-1}}$.

\noindent Sets I1 and I2, representing the outer region, are only extracted from the ACIS-I data, whereas Sets S1 and S2, representing the inner region, are only extracted from ACIS-S. 
Distinguishing the brighter and fainter sources in each region allows us to probe any difference in their mean spectral properties. 
It is noteworthy that we have excluded the 9 NIR counterparts and 11 LMXB transients (including CXOGC J174537.6-290035), since we are primarily concerned with the putative CVs.
The 412 foreground sources (Section~\ref{sec:hr}) are also excluded.

We extract the cumulative spectrum of a given set, using the CIAO tool {\it specextract}.
The source and background counts are extracted in the same regions as in the photometry procedure (Section~\ref{sec:pho}).
The ancillary response files (ARFs) and redistribution matrix files (RMFs) are obtained by weighting the individual source locations according to the local effective exposure.
The four spectra are shown in Figure \ref{fig:spec}.
The most obvious features in these spectra are the three Fe lines, although not all three lines are prominent in all spectra. 
Weaker lines can also be seen at 2–-2.5 keV, which are likely due to S and/or Si (see also \citealp{Muno2004}).
Therefore, we conduct spectral analysis using XSPEC v12.9.0n, over the energy range of 2.5--8 keV (Figure \ref{fig:spec}).
We note that neglecting energies below 2.5 keV does not affect the determination of the iron lines, which are of our primary concern.
We adopt a phenomenological model of bremsstrahlung plus three Gaussian lines centered around 6.4, 6.7 and 7.0 keV. 
These model components are subject to two absorption columns, one representing a partial covering absorption ({\it pcfabs} in XSPEC) that mimics local absorption in the putative CVs, and the other representing any additional absorption in the foreground ({\it wabs} in XSPEC).
The {\it wabs} absorption column is found to be well constrained in the brighter groups (S1 and I1), but is poorly constrained in the case of the fainter groups (S2 and I2); hence we fix its value at the best-fit value obtained from S1 and I1, respectively.

It turns out that the relatively hard continuum renders the plasma temperature ($T_{\rm b}$) not well constrained, hence we fix $T_{\rm b}$ at 40 keV, which is about the lower bound suggested by the data, but is typical of IPs when spectra up to a few tens of keV are available (e.g., \citealp{Xu2016}; \citealp{Hailey2016}). 
We verify that the line measurements are insensitive to the exact choice of $T_{\rm b}$.
We find that the line centroid and energy dispersion are generally well constrained when the line is significant; for spectrum I2, in which the 6.4 and 7.0 keV lines are absent or weak, we fix their centroid and dispersion at the best-fit values found in spectrum I1. 
The model provides a good fit to all four spectra.
We apply a bootstrapping method via {\it multifake} in XSPEC to derive the 68.3\% confidence range for the EWs and flux ratios of the 6.4 or 7.0 keV line to the 6.7 keV line ($I_{6.4}/I_{6.7}$ and $I_{7.0}/I_{6.7}$).  
The spectral fit results are summarized in Table \ref{tab:line}.

For comparison, we also analyze the cumulative spectrum of 232 LW sources with photon fluxes above $2\times10^{-7}{\rm~photon~cm^{-2}~s^{-1}}$ and located at $> 250\arcsec$ from the mean geometric center ($[RA,DEC]$ = [17:51:27.30, -29:35:05]), again to ensure an optimal spectral resolution. 
Although we have detected even fainter ($L_{\rm X} \lesssim 10^{31}{\rm~erg~s^{-1}}$) sources in the LW, such sources are thought to consist mainly of coronally active binaries (\citealp{Revnivtsev2009}), which have a different characteristic spectrum from that of CVs. Hence we do not include the fainter sources. 
As shown in Figure \ref{fig:spec_lw}, the LW spectrum has weak 6.4 and 7.0 keV lines with respect to the 6.7 keV line, which was hinted in previous work (\citealp{Morihana2013}) and is markedly different from the GC spectra (except perhaps I2).
The continuum of LW also appears softer than that of the GC sources, which is not a result of the lower line-of-sight column density ($N_{\rm H} \approx 10^{22}{\rm~cm^{-2}}$), but is rather dictated by a lower plasma temperature $T_{\rm b} \approx 18$ keV (Table \ref{tab:line}).  

Using the best-fit models of all four GC spectra, we compute the mean photon flux-to-energy flux conversion factor, $2.38\times10^{-8}{\rm~erg~photon^{-1}}$ (with only $\sim$10\% scatter), which has been applied to derive the unabsorbed 2--10 keV fluxes in Table~\ref{tab:main}.
Similarly, a conversion factor of $9.1\times10^{-9}{\rm~erg~photon^{-1}}$ is found for the LW sources.

\subsection{Flux Distribution}
\label{sec:LF}
Next, we explore the source flux distribution, or equivalently, the luminosity function, given nearly identical distances and similar spectra of the GC sources.
Again, we exclude the NIR counterparts and X-ray transients, which would otherwise contaminate the bright-end of the flux distribution.
Several regions of low detection sensitivity (hence a high degree of incompleteness), all at negative Galactic latitudes, are clearly seen in Figure~\ref{fig:2d}, which are due to the presence of Sgr A East and two molecular clouds (Mezger et al.~1996).     
Therefore, we further exclude from the following analysis sources of $\delta b <0$ (relative to Sgr A*). 
An intrinsic symmetry about the Galactic plane can be expected for sources in the NSC, hence results from this choice should be statistically meaningful.  

The observed flux distributions for the inner ($R < 250\arcsec$) and outer ($250\arcsec < R < 500\arcsec$) regions are shown in Figure~\ref{fig:LF}, which involve 649 and 1511 sources, respectively. 
In Figure~\ref{fig:LF}a, the flux distributions are normalized by the underlying sky area, while in Figure~\ref{fig:LF}b normalized by the projected enclosed stellar mass in each region, as derived in Section~\ref{sec:spatial}.
The two distributions have similar amplitudes at the bright-end, rise steeply to $S_{\rm 2-8} \approx$ a few $10^{-7}{\rm~photon~cm^{-2}~s^{-1}}$, and turn over toward lower fluxes, which is a joint effect of detection incompleteness and the Eddington bias.
We fit the flux distributions using a canonical power-law model, $N (> S) = K_{\rm s} (S/S_0)^{-\alpha}$. The scaling factor $K_{\rm s}$ is normalized by enclosed stellar mass; the nominal flux $S_0$ is set to $10^{-7}$ photon~cm$^{-2}$~s$^{-1}$.    
This model is corrected for the incompleteness and Eddington bias, following the procedure of \citet{Wang2004} and \citet{Li2010} and taking into account the source spatial distribution, which to first order should follow the NIR starlight (Section~\ref{sec:spatial}).
To minimize systematics at the lowest fluxes, we restrict the fit to sources having fluxes above the 50\% completeness in each region (Figure~\ref{fig:cm}).
We also account for contribution by the cosmic X-ray background (CXB), 
adopting the empirical $logN-logS$ relation of \citet{Georgakakis2008} and assuming for the CXB an intrinsic power-law spectrum with a photon-index of 1.4 and an absorption column density of $2\times10^{23}{\rm~cm^{-2}}$, i.e., assuming twice the line-of-sight extinction to the GC.
As shown by the solid curves in Figure~\ref{fig:LF}a, the CXB component is everywhere almost two orders of magnitude lower than the GC sources. 
The total expected number of CXB sources is only 11 and 10 in the inner and outer regions, respectively.
By minimizing the $C$-statistic (Cash 1979), we obtain $\alpha=1.63^{+0.16}_{-0.15}$ ($1.65^{+0.10}_{-0.09}$) and $K_{\rm s} = 1.37^{+0.67}_{-0.63}~(1.56^{+0.24}_{-0.20}) \times10^{-4}{\rm~M_\odot^{-1}}$ for the inner (outer) region, i.e., the two flux distributions are statistically consistent with each other.
At face value, the slope is also consistent with $\alpha = 1.5\pm0.1$ found by M09 for their sources detected within the inner 8$\arcmin$. 
Nevertheless, our result is based on a more careful treatment of systematics and extends to much lower fluxes. 

For comparison, we also construct the flux distribution of the LW sources, selecting 324 sources detected within $250\arcsec$ from the mean geometric center, this time to ensure an optimal sensitivity. 
We note that the CXB component, with an expected number of 23 sources, is also minor in the LW. To obtain this estimate, we have similarly assumed the $logN-logS$ relation of the CXB and a column density of $1.4\times10^{22}{\rm~cm^{-2}}$, i.e., twice the line-of-sight value to the LW.
We fit the flux distribution down to $\sim$$2\times10^{-7}{\rm~photon~cm^{-2}~s^{-1}}$, obtaining $\alpha = 0.82^{+0.07}_{-0.06}$ 
and $K_{\rm s} = 1.30^{+0.12}_{-0.12}\times10^{-4}{\rm~M_\odot^{-1}}$, based on a surface stellar mass density of $ 4.6\times 10^{4}{\rm~M_\odot~arcmin^{-2}}$ (\citealp{Revnivtsev2010}).
Clearly, the luminosity function over $L_{\rm X} \approx 10^{31-33}{\rm~erg~s^{-1}}$ is significantly steeper in the GC than in the LW (Figure~\ref{fig:LF}). 
We emphasize that this a robust result, regardless of the adopted model.
Along with the distinct spectra (Section~\ref{sec:spec}), this strongly suggests different X-ray source populations in the nuclear bulge and the Galactic bulge. 

\subsection{Spatial Distribution}
\label{sec:spatial}
We now turn to the spatial distribution of the X-ray sources.
\citet{Muno2003} studied the surface density profile of sources in the inner 8$\arcmin$ with $S_{2-8} > 5\times10^{-7}{\rm~photon~cm^{-2}~s^{-1}}$, finding that it roughly follows an $R^{-1}$ trend.
M09 further examined the surface density profile within the 2$\degree \times0.8 \degree$ field, but only for sources with fluxes above $2\times10^{-6}{\rm~photon~cm^{-2}~s^{-1}}$.
They found that the profile agrees well with the projected stellar mass distribution measured from NIR observations \citep{Launhardt2002}, when the latter is normalized by an X-ray source abundance of $5\times10^{-7}{\rm~M_\odot^{-1}}$.

We revisit the azimuthally-averaged surface density profile (Figure~\ref{fig:spatial}), selecting only 779 sources with $\delta b >0$ and $S_{2-8} > 3.5\times10^{-7}{\rm~photon~cm^{-2}~s^{-1}}$, i.e., approximately the 90\% completeness limit for $R < 250\arcsec$ (Section~\ref{sec:complete}).  
To compare with the stellar mass distribution in the nuclear bulge, 
we adopt the latest modeled distribution given by \citet{Fritz2016}, which is based on a systematic analysis of a large set of NIR observations. 
The model consists of two morphologically distinct components of the nuclear bulge: an inner component representing the NSC, and an outer component tracing the {\it nuclear disk}  \citep{Launhardt2002}.
Each component is parameterized by the $\gamma$-model (Dehnen 1993), 
\begin{equation}
\rho_{\rm s}(r) = \frac{3-\gamma}{4\pi}  \frac{L}{r^{\gamma}} \frac{a}{(r+a)^{4-\gamma}},
\end{equation}
where $r$ is the spherical radius. We have adopted the following parameters:
$L_{\rm in}$ = $6.73\times10^{4}$ stars, $a_{\rm in}$=194$\arcsec$, $\gamma_{\rm in}$ = 0.90, $L_{\rm out}$ = $7.05\times10^{6}$ stars, $a_{\rm out}$ = 3396$\arcsec$, $\gamma_{\rm out}$ = 0, which provide a good characterization of the radial density distribution of resolved giant stars in the nuclear bulge.
The cumulative mass of the two components within $r = 100\arcsec$ was determined to be
$(6.09 \pm 0.97) \times 10^{6}{\rm~M_{\odot}}$ (\citealp{Fritz2016}), assuming a constant mass-to-light ratio.
The surface mass density then follows,
\begin{equation}
\Sigma(R) = 2 \int_{R}^{\infty}~\rho_{\rm s}(r)rdr/\sqrt{r^{2}-R^{2}}.
\end{equation}
The NSC dominates the surface density within a projected radius of $R \approx 200\arcsec$, outside of which the nuclear disk dominates.
We find that the surface density profile can be well characterized by the above mass model (solid curve in Figure~\ref{fig:spatial}), provided that the X-ray source abundance is $(1.76\pm0.10)\times10^{-5}{\rm~M_{\odot}^{-1}}$ for $S_{2-8} > 3.5\times10^{-7}{\rm~photon~cm^{-2}~s^{-1}}$. 
We note that the outermost data point ($R \gtrsim 400\arcsec$) in the profile lies slightly below the modeled curve, due to a lower degree of completeness at the flux threshold. 
On the other hand, there is no source detected within the innermost $3\farcs5$ (after excluding the transient CXOGC\,J174540.0-290031), where the mass distribution peaks. 
This central stellar ``cusp" predicts only $\sim$1 source (above the adopted flux threshold), hence the apparent deficiency within the central $3\farcs5$ is statistically insignificant (see Section~\ref{sec:sum} for further discussion).
 
Naturally, the above fitted source abundance is consistent with the value of $K_{\rm s}$ obtained in Section~\ref{sec:LF}, when extrapolated to $S_{2-8} = 1.0\times10^{-7}{\rm~photon~cm^{-2}~s^{-1}}$.
We can further translate this number to $\sim$$1.0\times10^{-6}{\rm~M_{\odot}^{-1}}$ for $S_{2-8} > 2\times10^{-6}{\rm~photon~cm^{-2}~s^{-1}}$, which would then be a factor of 2 higher than the abundance found by M09 in the 2$\degree \times0.8 \degree$ field.  
Indeed, M09 recognized an excess in the observed surface density within the inner few arcminutes, after matching their profile at larger radii with $\sim$2 times lower abundance. 
We now show that what appeared an ``excess" in M09 actually traces the underlying stellar distribution in the NSC. 

The above analysis has assumed spherical symmetry. As shown by \citet{Fritz2016}, the nuclear bulge in NIR appears more flattened toward larger projected radii, due to the increasing contribution from the nuclear disk. For comparison, we explore the two-dimensional X-ray source distribution, assisted with the CIAO tool {\it Sherpa}.
We fit the same set of sources as used in the radial profile, now binned into 10$\arcsec$-wide grids. Similar to \citet{Fritz2016}, we adopt a 2-d S\'ersic model, forcing the minor-axis to be aligned with Galactic latitude. This results in an ellipticity of $0.40^{+0.13}_{-0.13}$,
which is in rough agreement with the degree of flattening over a similar radial range found in Fritz et al.~(2016; Table 3 therein).




\section{Discussion and Summary} {\label{sec:sum}
We have presented an updated catalog of X-ray point sources in the inner 500$\arcsec$ ($\sim$20 pc) of the GC, based on 1.5 Ms ACIS-I and 2.94 Ms ACIS-S observations. This ultra-deep dataset allows us to achieve a factor of $\sim$2 better sensitivity for probing faint X-ray sources in the NSC, compared to the last major source catalog of M09.
We have detected 3540 sources that are probably located in the GC,
down to a 2--10 keV unabsorbed luminosity of $\sim$$1.0\times10^{31}{\rm~erg~s^{-1}}$,
among which $\sim$1300 are new identifications.
Meanwhile, $\sim$1100 sources cataloged in M09 have no apparent counterpart in our catalog, which might be largely attributed to background fluctuations of relatively low significance as appearing in the much shallower data of M09.

Since the first detection of faint X-ray sources ($L_{\rm X} \approx 10^{31-34}{\rm~erg~s^{-1}}$) in the GC (Wang et al.~2002; Muno et al.~2003), understanding the very nature of these sources remains a key unresolved issue. At present, the leading candidate for the responsible population is IPs, a subclass of mCVs, for reasons mentioned in Section~\ref{sec:intro}. 
This scenario, however, can be challenged by two legitimate questions: (i) Can quiescent LMXBs (qLMXBs) have a substantial contribution to the observed source populations? 
It has long been suggested that BHs, and on a longer timescale NSs, would migrate into the NSC, due to dynamical friction over the background of low-mass stars (Morris 1993). The frequent star formation in the NSC over the past few hundred Myrs (e.g., Pfuhl et al.~2011) could be another source of BHs and NSs. The number of BHs and NSs currently residing in the NSC is highly uncertain, but is estimated to be at least a few $10^4$ (e.g., Deegan \& Nayakshin 2007). If a significant fraction of these compact remnants were eventually locked up in close binaries, which is plausible in the dense stellar environment of the NSC, they would spend most of the time manifesting themselves as qLMXBs with $L_{\rm X} \approx 10^{31-34}{\rm~erg~s^{-1}}$.  
(ii) Even if qLMXBs are a negligible population, can non-magnetic CVs, in particular dwarf novae (DNe), make a substantial contribution? Generally speaking, IPs tend to be more X-ray ($\gtrsim$ 2 keV) luminous and exhibit stronger Fe 7.0 keV and 6.4 keV lines (with respect to the 6.7 keV line) and a harder continuum, as compared to DNe, chiefly due to a shock produced along the accretion column towards the magnetic poles (see recent review by Mukai 2017).
However, individual DNe can also show these characteristics in their X-ray emission, especially when a large WD mass (hence a smaller WD radius and deeper gravitational potential) is involved. 
Moreover, the intrinsic fraction of mCVs among all CVs in the field is estimated to be on the order of 10\% (Mukai 2017). If most of the faint X-ray sources in the NSC are indeed IPs, one may infer that a much larger number of non-magnetic CVs awaits discovery at still fainter fluxes.
Below, we address these two questions, assisted with the new insights on the global source properties obtained in Section~\ref{sec:global}. 

Figure~\ref{fig:lineratio} displays EW$_{6.7}$ or $I_{6.4}/I_{6.7}$ versus $I_{7.0}/I_{6.7}$ measured from the GC and LW spectra (Table~\ref{tab:line}), in comparison with a sample of field CVs studied by Xu et al.~(2016). 
This sample includes 17 IPs, 3 polars, 20 DNe and 8 symbiotic stars, but is not expected to be statistically representative for any subclass.
Nevertheless, from this sample a general trend is evident that IPs tend to have higher $I_{7.0}/I_{6.7}$, higher $I_{6.4}/I_{6.7}$ and smaller EW$_{6.7}$, when compared to the DNe. 
Interestingly, the GC and LW measurements seem to follow these trends. 
In particular, the S1 and I1 sources (filled and open circles in Figure~\ref{fig:lineratio}) have $I_{7.0}/I_{6.7}$ close to the mean value of field IPs ($0.71\pm0.04$; Xu et al.~2016), whereas the I2 sources (filled downward triangle) are more similar to DNe in $I_{7.0}/I_{6.7}$. The S2 sources (open downward triangle) appear peculiar in the diagrams, in the sense that they exhibit $I_{7.0}/I_{6.7}$ similar to IPs but with EW$_{6.7}$ and $I_{6.4}/I_{6.7}$ closer to DNe.    
A straightforward interpretation of Figure~\ref{fig:lineratio} is that the observed flux ratios and EW$_{6.7}$ reflect different fractions of constituent mCVs and non-magnetic CVs.
In this regard, I1 and S1 should be dominated by IPs, which is not surprising because both sets are composed of the brighter sources, i.e., with $L_{\rm X} \gtrsim 6\times10^{31}{\rm~erg~s^{-1}}$, a practical threshold above which nearly all the IPs in Xu et al.~are found.
On the other hand, a larger, perhaps predominant, fraction of the S2 and I2 sources should be DNe. 
The LW sources have flux ratios intermediate between I1/S1 and S2/I2, which at face value suggests comparable contributions from IPs and DNe. 
This is reasonable, since we have not distinguished brighter and fainter groups in the LW due to a smaller total number of sources.  
The above interpretation of the Fe lines is also indicative of a minor contribution by qLMXBs.
Empirically, quiescent BH-LMXBs exhibit a power-law X-ray spectrum without a significant Fe 6.7 keV line, which is understood as being dominated by jet synchrotron emission (e.g., Froning et al.~2016; Plotkin et al.~2016, 2017); the X-ray spectra of quiescent NS-LMXBs, often better explained by bremsstrahlung, are also typically free of the 6.7 keV line (Chakrabarty et al.~2014).
Therefore, if qLMXBs had a significant contribution to the detected X-ray sources in the NSC, they would simply dilute EW$_{6.7}$, which is not observed in any set of the spectra. 
Further improvement on the spectral decomposition invites a joint analysis of {\it Chandra} and NuSTAR data on the NSC. 
The latter (Perez et al.~2015), in particular, is crucial for tightly constraining the hard continuum ($\gtrsim$20 keV) presumably dominated by the IPs.

The NSC sources show a steep luminosity function over the range of $10^{31-33}{\rm~erg~s^{-1}}$ (Section~\ref{sec:LF}). 
We estimate the resolved fraction of the total (point sources plus diffuse) flux from the region of $R < 250\arcsec$ and ${\delta}b > 0$.
To do so, we first account for the instrumental background, using the ACIS ``stowed background" data\footnote{http://http://cxc.harvard.edu/contrib/maxim/stowed/}, to derive the genuine X-ray flux from the region. 
We then estimate the resolved fraction to be $\sim$20\% ($\sim$23\%) over the 2--8 (6.5--7.1) keV range.
Further corrected for the detection incompleteness down to $L_{\rm X} \approx 10^{31}{\rm~erg~s^{-1}}$, the resolved fraction becomes 24\% over 2--8 keV. 
This in turn requires that the luminosity function of the NSC become significantly flattened at $L_{\rm X} \lesssim 5\times10^{30}{\rm~erg~s^{-1}}$, to leave room for the truly diffuse emission that is clearly present in this region (e.g., Baganoff et al.~2003).
This can be tested, although much deeper {\it Chandra} data would be required.

Also enlightening is our finding that the spatial distribution of the X-ray sources agrees well with the NIR starlight distribution, which itself is a good proxy of the stellar mass distribution in the NSC, down to a scale of just $\sim$10$\arcsec$ (Section~\ref{sec:spatial}).  
This is not so trivial as it might appear. 
Regardless of the exact nature of the X-ray sources, they are no doubt close binary systems, the formation and evolution of which are readily affected by dynamical processes, such as single-single capture and binary-single exchanges (Hills 1975; Hut 1993). 
Indeed, the NSC, with a stellar density $\gtrsim 10^4{\rm~pc^{-3}}$, defines one of the densest environments in our Galaxy, which favors stellar encounters. 
Moreover, Sgr A* and a close binary located within its gravitational influence radius, $r_{\rm GI} \approx 1.5(M_{\rm BH}/4\times10^6{\rm~M_\odot})(\sigma_{\rm v}/150{\rm~km~s^{-1}})^{-2}{\rm~pc}$ ($\sigma_{\rm v}$ the stellar velocity dispersion; Fritz et al.~2016), would form a hierarchical triple system. 
Under this circumstance, the binary would be subject to the Kozai-Lidov effect (see review by Naoz 2016), possibly resulting in a highly elliptical orbit that could lead to a merger of the two stars. 
This might be relevant to the absence of X-ray sources in the innermost $3\farcs5$ (Figure~\ref{fig:spatial}), although the deficiency there is statistically insignificant. 
The fact that the radial distribution of faint X-ray sources closely follows the NIR starlight thus suggests that dynamical effects leading to the formation and disruption of CVs in the NSC are either inefficient or canceling each other.

On the other hand, one may expect that the radial distribution of LMXBs is more concentrated than that of the CVs, 
due to mass segregation that radially separates BHs and NSs from the lighter WDs (Morris 1993).
Indeed, an over-abundance of X-ray transients (presumably LMXBs) in the central parsec has been pointed out by Muno et al.~(2005). 
To update this view, we plot the surface density profile of currently known X-ray transients as orange points in Figure~\ref{fig:spatial}. The much steeper profile of the transients, compared to the faint X-ray sources, is remarkable.   
Since it is unlikely that many more transients have been missed by existing X-ray observations of the NSC (see below), it is reasonable to assume that the known transients, with an abundance of $\sim$$6\times10^{-7}{\rm~M_\odot^{-1}}$, represent the entire qLMXB population in terms of their spatial distribution in the NSC. 
We further compare the observed radial distribution of the transients with a projected $\rho_s^2(r)$ (Equation 4) distribution, shown as an orange curve in Figure~\ref{fig:spatial}, motivated by the expectation that most LMXBs in the NSC were dynamically-formed, i.e., essentially through stellar encounters such as tidal captures, direct collisions and binary-single exchanges.
It is intriguing that the observed distribution can be roughly matched by the modeled distribution. 
We point out that Voss \& Gilfanov~(2007) reported a similar behavior of the luminous ($L_{\rm X} \gtrsim 10^{36}{\rm~erg~s^{-1}}$) LMXBs detected in the inner $\sim$200 pc of M\,31, which they attributed to dynamical formation.
The markedly different radial distributions between the transients and the faint X-ray sources again strongly suggests that qLMXBs cannot be a significant population in the NSC.  
In fact, just $\sim$30 hidden qLMXBs following the $\rho_s^2(r)$ distribution would have caused a notable deviation between the observed and modeled (i.e., projected $\rho_s$) distributions of the faint sources in the inner 35$\arcsec$. 
This in turn constrains the relative contribution of qLMXBs to $\lesssim$5\% of the detected sources. It is noteworthy that both BH-LMXBs and NS-LMXBs in extreme quiescence can have X-ray luminosities below our current detection limit of $\sim$$10^{31}{\rm~erg~s^{-1}}$ (\citealp{Garcia2001}; \citealp{Heinke2010}). 
Such extreme qLMXBs, as well as isolated BHs and NSs, remain to be explored in the NSC.

Our final remarks are for the mild variability of most detected sources, on timescales from days up to 13.5 yrs (Section~\ref{sec:var}).
This can also be understood if these sources are primarily CVs. 
Among the known classes of X-ray sources with quiescent luminosities of $10^{31}{\rm~erg~s^{-1}} \lesssim L_{\rm X} \lesssim 10^{34}{\rm~erg~s^{-1}}$, qLMXBs are the best candidates to become transients which increase their fluxes by a factor of $\gtrsim100$ over a timescale of weeks to months.
Indeed, the {\it Swift} monitoring program had captured about a dozen outbursting LMXBs, after observing the NSC for a period of 8--9 months annually during 2006--2014, at a cadence of once every 1--10 days (Degenaar et al.~2015).   
It is highly unlikely that significantly more LMXBs in their outbursting phase, which typically lasts for weeks, had been missed. 
On the other hand, empirically assuming that an LMXB gives rise to an outburst once in 50--100 yrs,
finding $\sim$10 transients in the past decade thus suggests that no more than $\sim$150 qLMXBs exist among the $\sim$3000 faint X-ray sources. 
Such a small fraction is again compatible with the above constraints imposed by the cumulative spectra and spatial distribution. 

\acknowledgements
This work is supported by National Science Foundation of China under grants 11473010 and 11133001. 
We thank Bin Luo for helpful suggestions on the catalog, and thank Xiang-Dong Li, Smadar Naoz, Rainer Sch\"{o}del, Xiao-Jie Xu and Shuo Zhang for useful discussions. 
M.M. acknowledges support from SAO/NASA grant 16151X for research with {\it Chandra}.
Z.Z. and Z.L. are grateful to the hospitality of UCLA during their visits. 
Z.L. acknowledges support from the Recruitment Program of Global Youth Experts.

\begin{figure}
      \centering
        \includegraphics[scale=0.7]{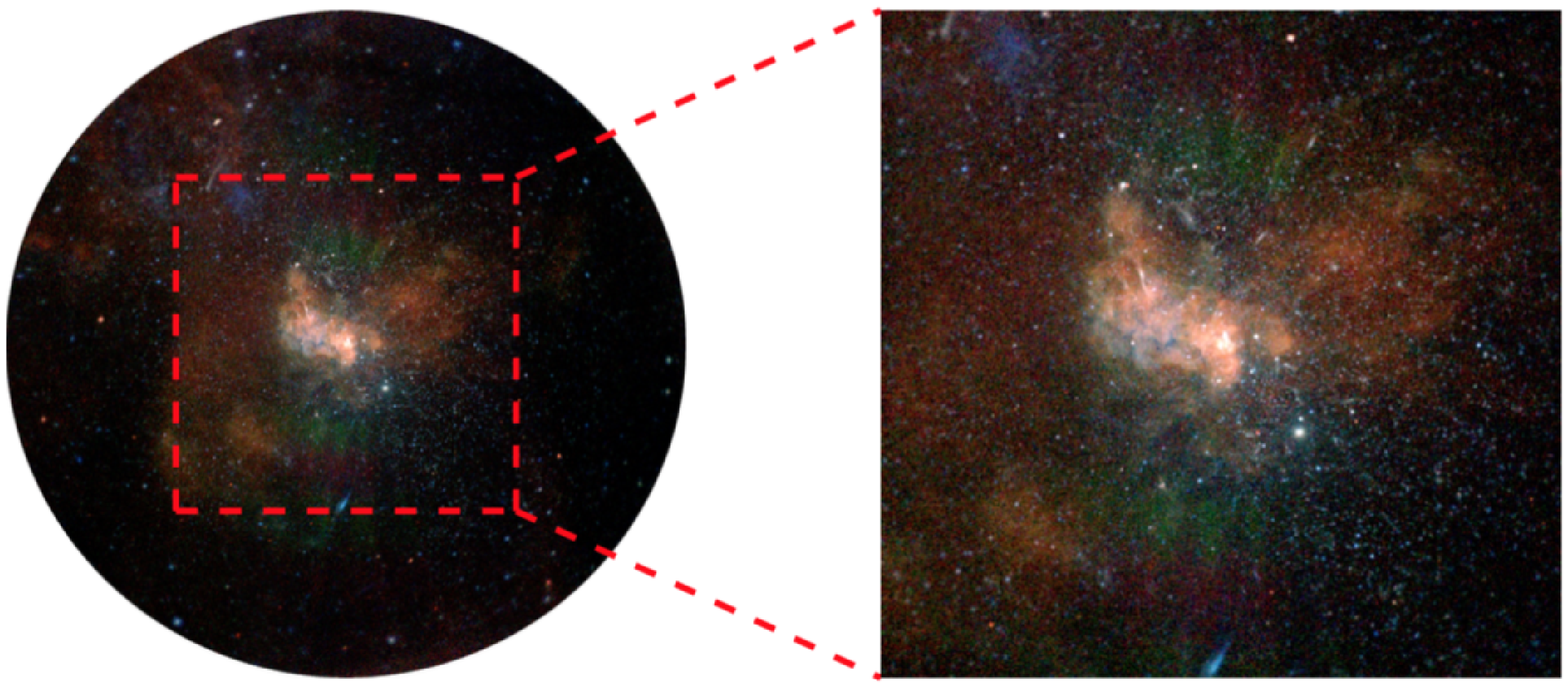}
       \caption{Tri-color image of the inner 500$\arcsec$ region ($\sim$20 pc) of the Galactic center, as seen by the 4.44 Ms {\it Chandra}/ACIS observations. Red for 2--3.3 keV, green for 3.3--4.7 keV and blue for 4.7--8 keV. The intensity in each band has been corrected for effective exposure and smoothed with a Gaussian kernel of 2 pixels. 
The zoom-in panel has a dimension of $250\arcsec \times 250\arcsec$.
The greenish regions are artifacts due to first-order dispersed photons produced in the ACIS-S/HETG observations, which have little effect in the source detection and characterization.
}
     \label{fig:tricolor}
\end{figure}

\begin{figure}
	\centering
       \includegraphics[scale=0.9]{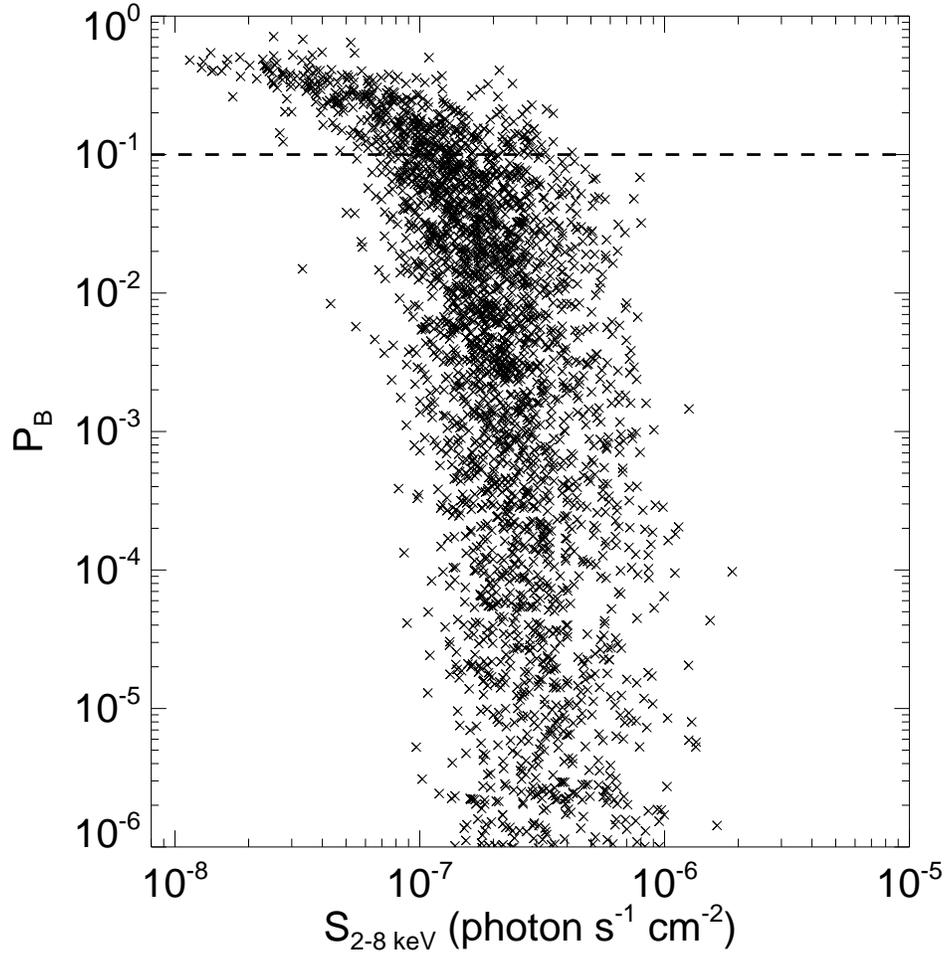}
	\caption{Binomial no-source probability as a function of the 2--8 keV photon flux. The dashed line marks the threshold ($P_{\rm B} > 90\%$) for filtering spurious detections, i.e., due to background fluctuations.}
	\label{fig:pb}
\end{figure}

\begin{figure}
	\centering
       \includegraphics[scale=1]{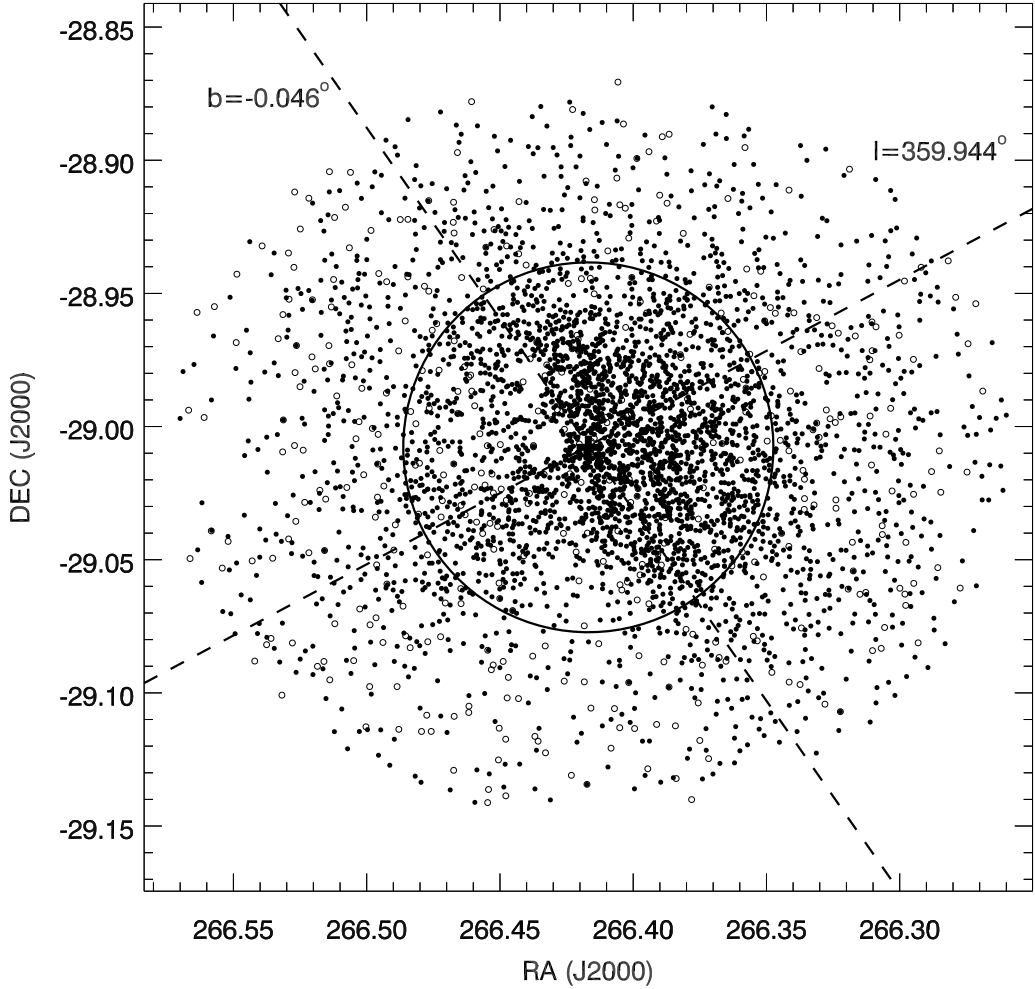}
	\caption{Spatial distribution of the detected sources within the inner 500$\arcsec$. 
The Galactic longitude and latitude axes passing through Sgr A* are plotted with dashed lines. 
Open circles denote foreground sources, while sources located in or beyond the GC are marked with dots. 
The large circle outlines a projected radius of 250$\arcsec$, within which sources are covered by all 87 ACIS observations. 
At $b < -0.046$, there are three notable regions in deficit of sources, which is due to the presence of the Sgr A East SNR, and two molecular clouds, M-0.02-0.07 and M-0.13-0.08 (\citealp{Mezger1996}).
}
	\label{fig:2d}
\end{figure}

\begin{figure}
        \centering
       \includegraphics[width=0.8\textwidth]{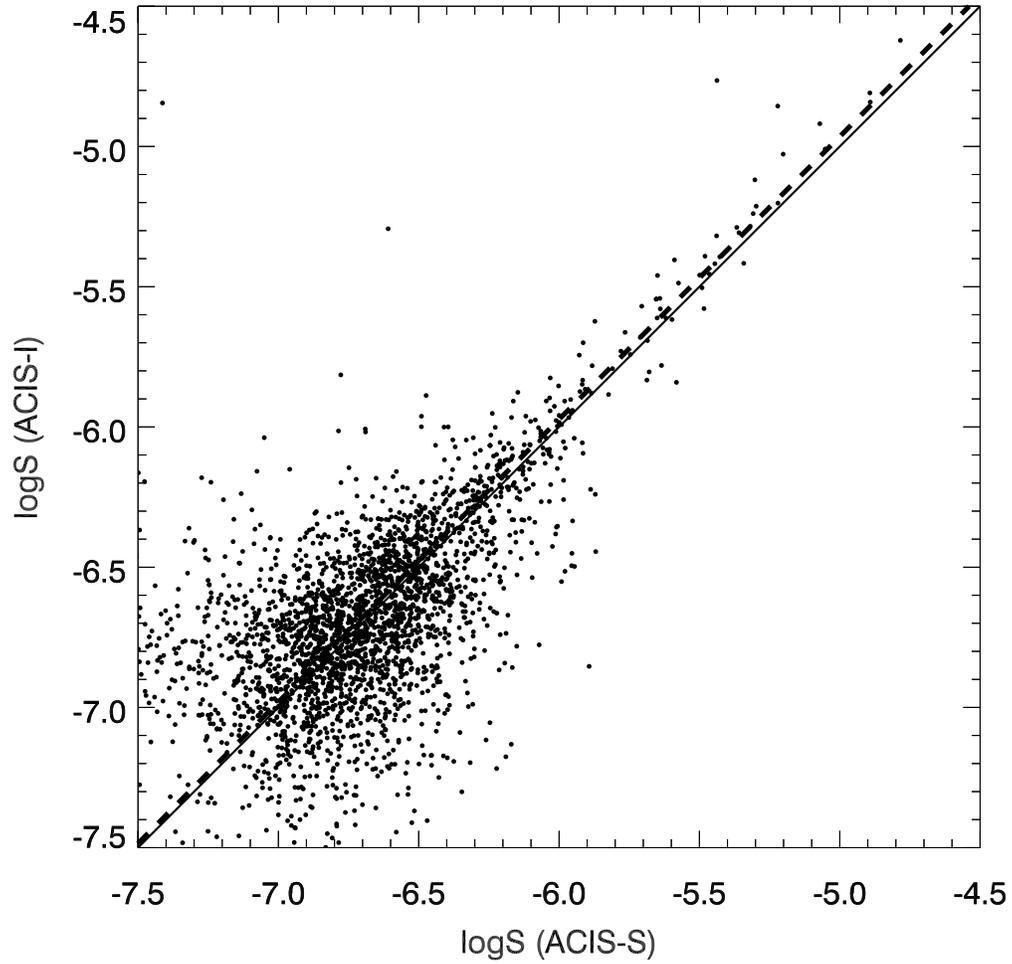}
	\caption{Comparison of the 2--8 keV photon fluxes of common sources measured with ACIS-I data and ACIS-S data.  The diagonal solid line shows a 1:1 correspondence. The dashed line represents a linear fit, $y = (1.008\pm0.005) x + (0.075\pm0.032)$, indicating no significant bias between the ACIS-I and ACIS-S measurements.}
	\label{fig:is}
\end{figure}

\begin{figure}
	\centering
	        \includegraphics[scale=0.45]{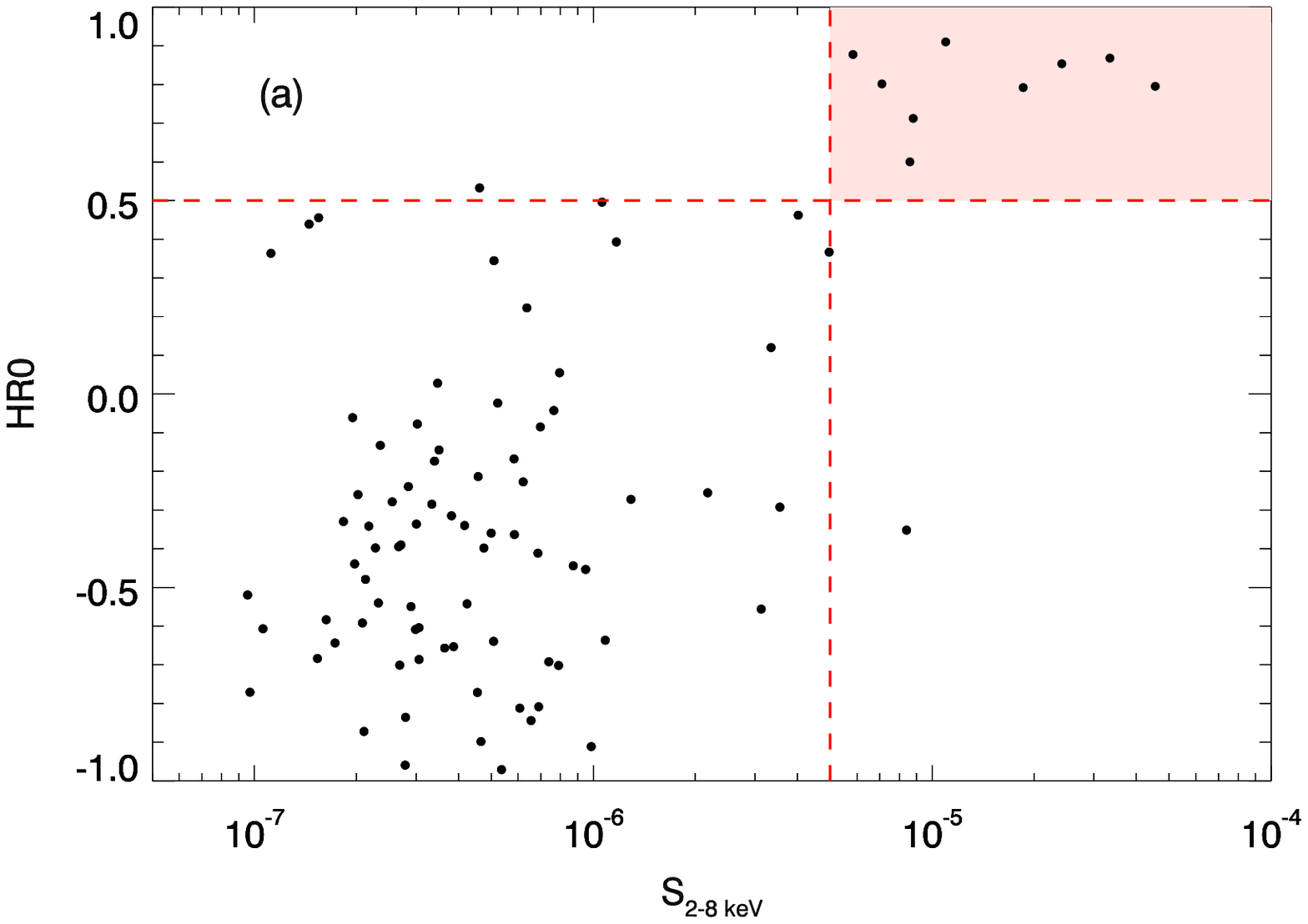}
	        \hspace{5pt}
		\includegraphics[scale=0.45]{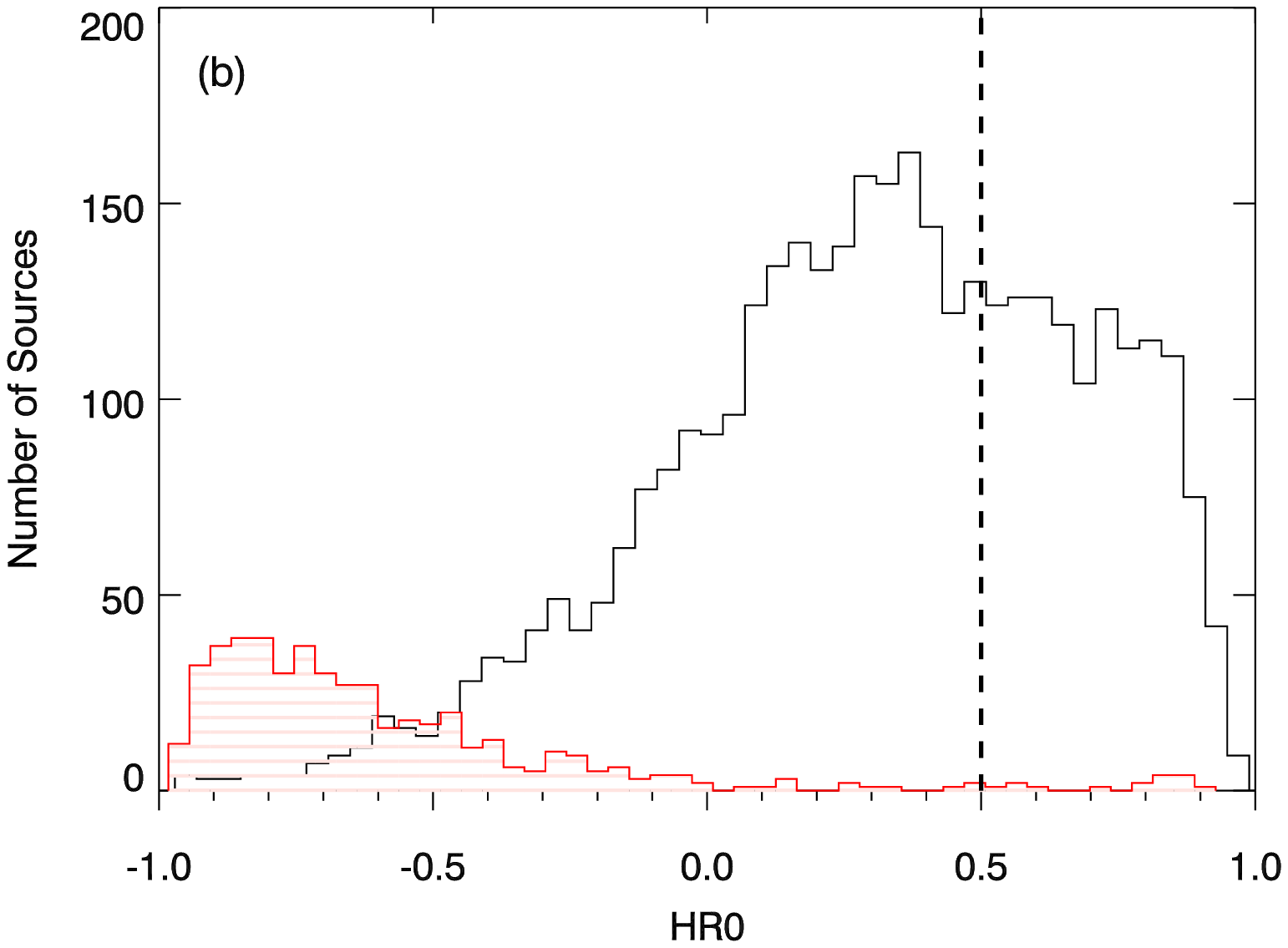}
		\\[5pt]
			\includegraphics[scale=0.45]{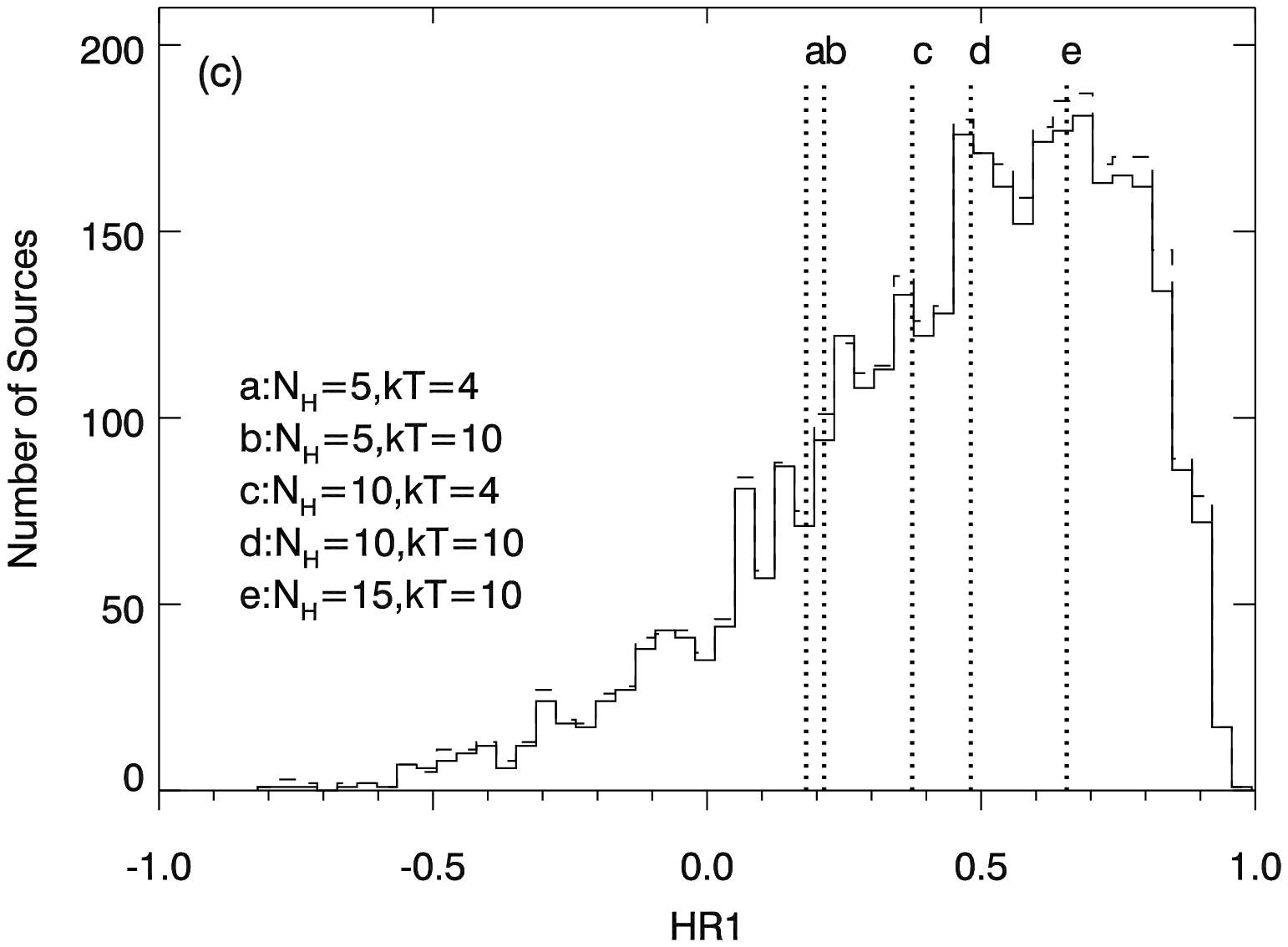}
			        \hspace{10pt}
			\includegraphics[scale=0.45]{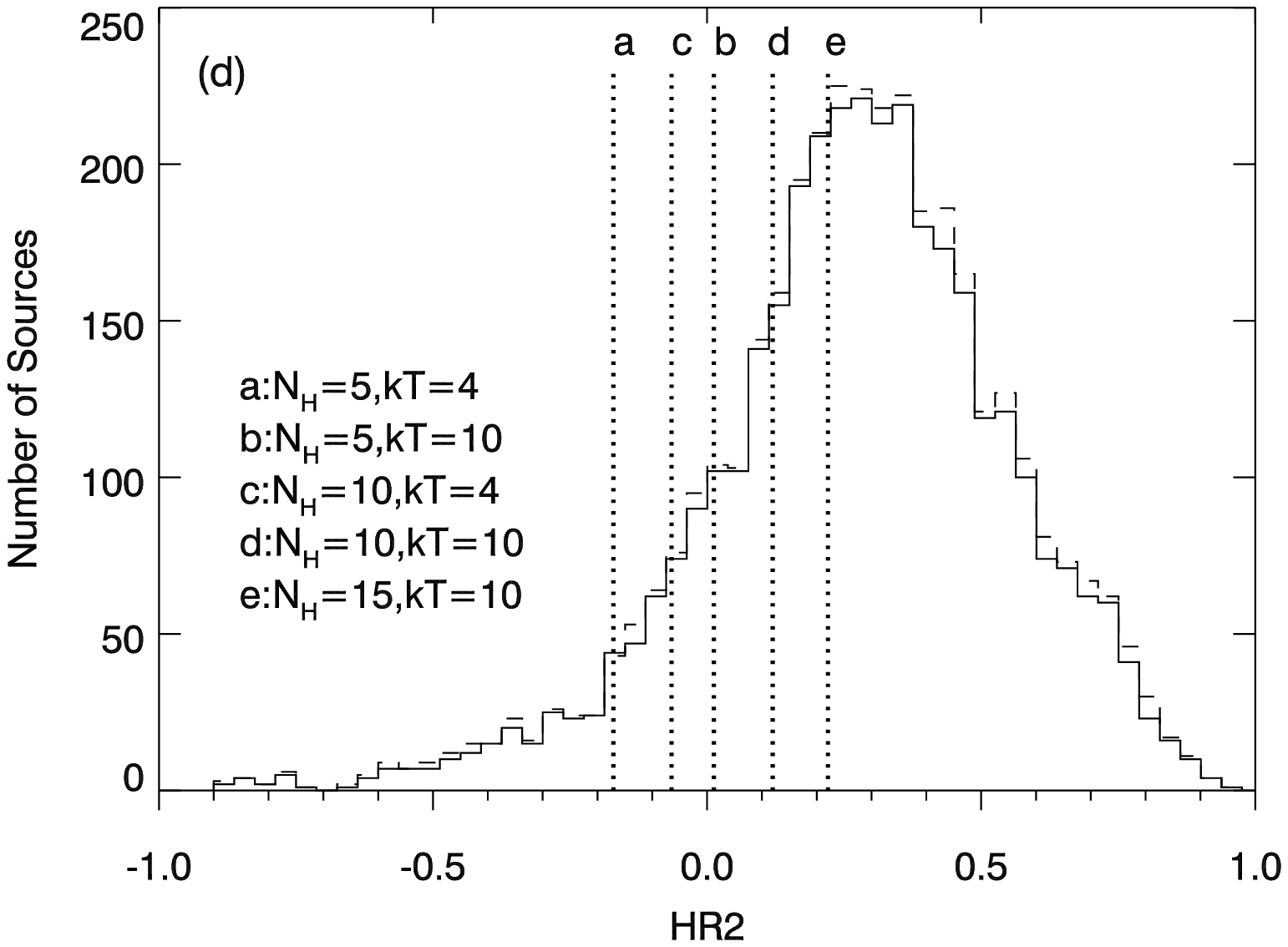}

	\caption{Histograms of the measured hardness ratios (see Section~\ref{sec:hr} for definition). (a) Sources detected in both the 0.5--2 and 2--8 keV bands. Those found in the upper right corner, with HR0 $> 0.5$ and $S_{2-8} > 5\times10^{-6}{\rm~photon~cm^{-2}~s^{-1}}$, are considered residing in the GC. 
(b) The black and red curves are for sources detected in 2--8 keV and 0.5--2 keV, respectively. 
(c) HR1 and (d) HR2. 
The vertical dotted lines denote hardness ratios predicted by various absorbed bremsstrahlung models. 
}
	\label{fig:hr}
\end{figure}

\begin{figure}
        \centering
         \includegraphics[scale=0.7]{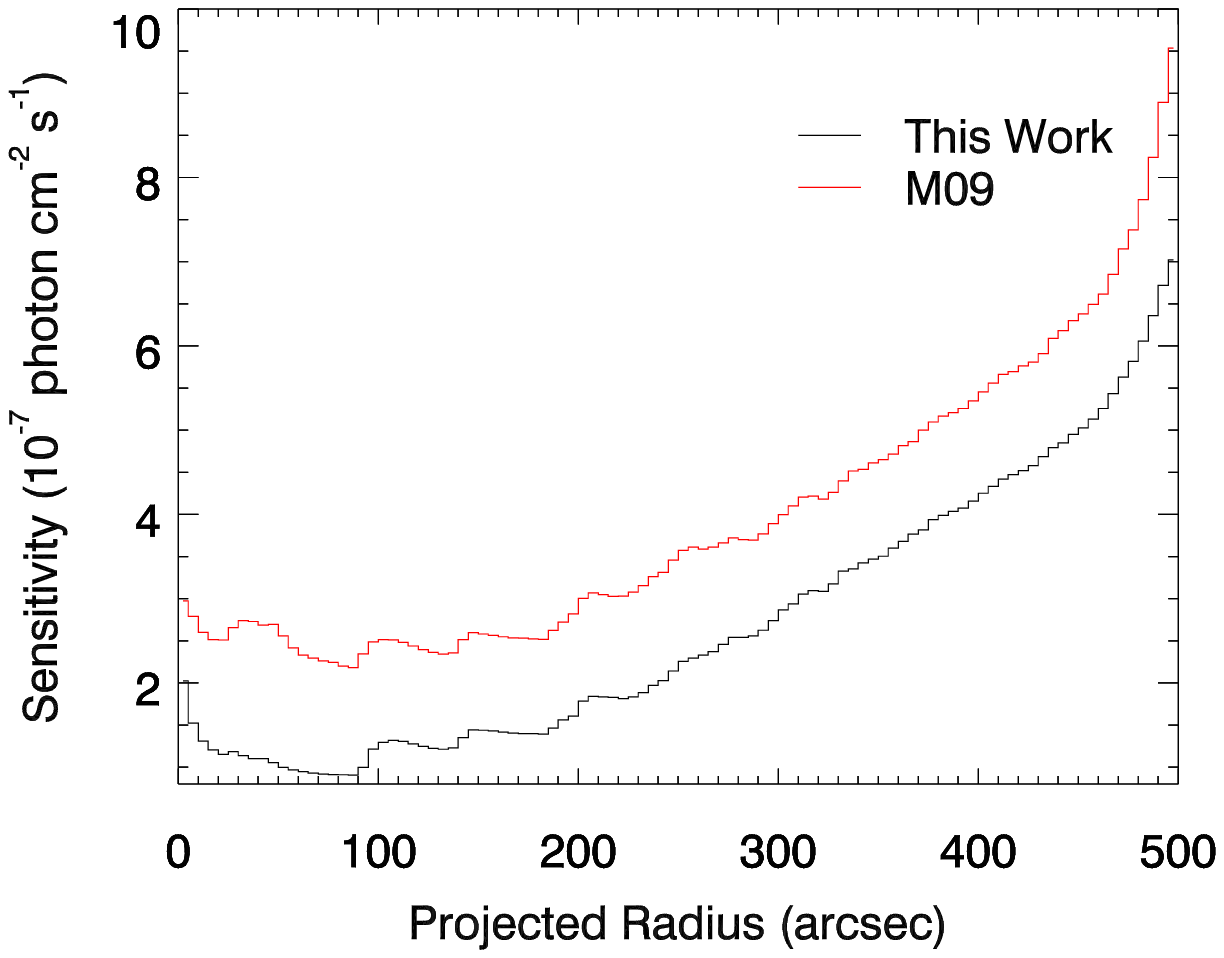}
               \\[20pt]
        \includegraphics[scale=0.7]{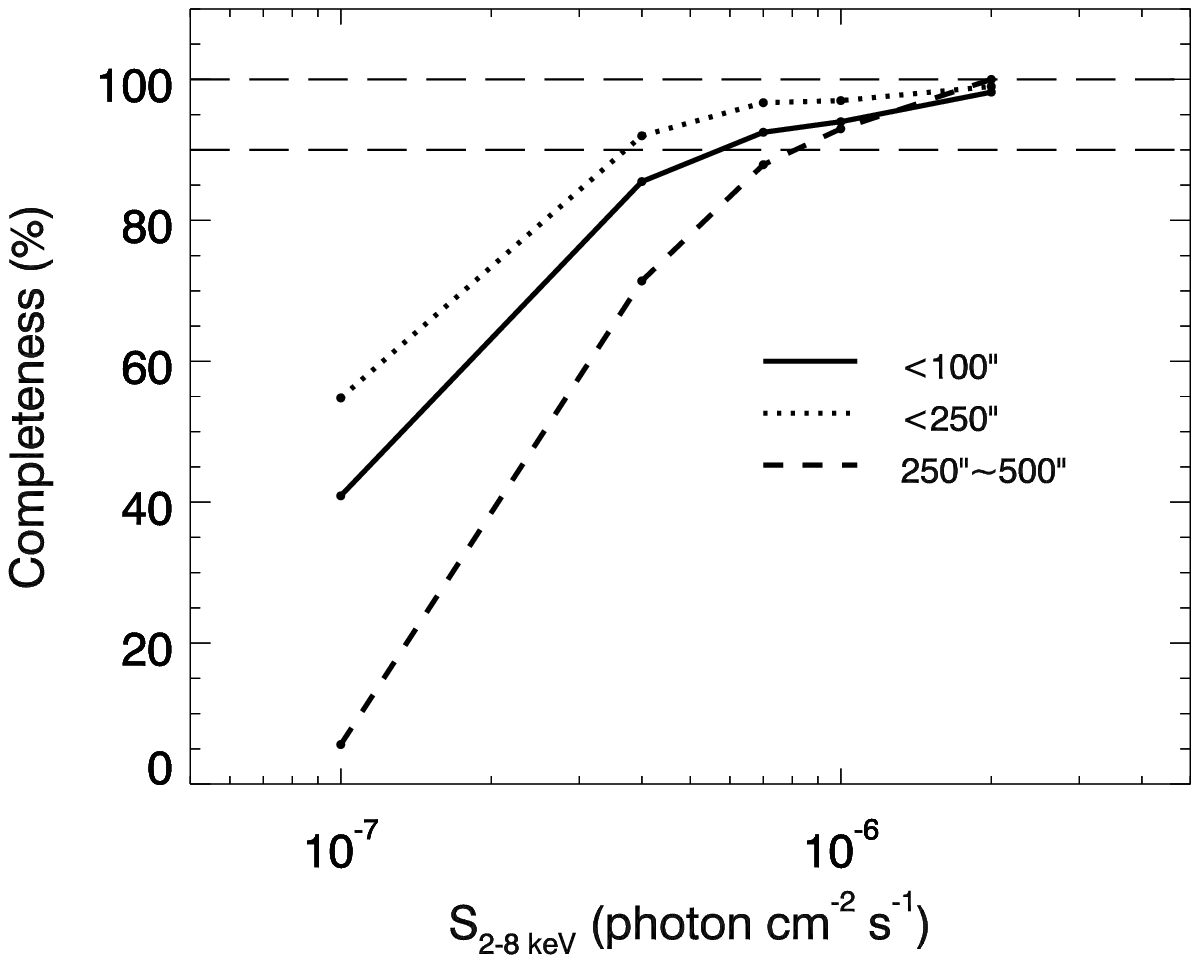}
	\caption{{\it Top panel}: Median sensitivity as a function of projected radius. The black and red curves represent the sensitivity achieved in this work and that would have been achieved by M09 if an identical detection procedure had been adopted, respectively. 
{\it Bottom panel}: Detection completeness as a function of observed photon flux. The solid, dotted and dashed lines denote the curves for sources located at $R < 100\arcsec$, $R < 250\arcsec$ and $250\arcsec < R < 500\arcsec$, respectively. 
The two horizontal long dashed lines mark completeness levels of 90\% and 100\%.}
	\label{fig:cm}
\end{figure}

\begin{figure}
        \centering
       \includegraphics[scale=0.5]{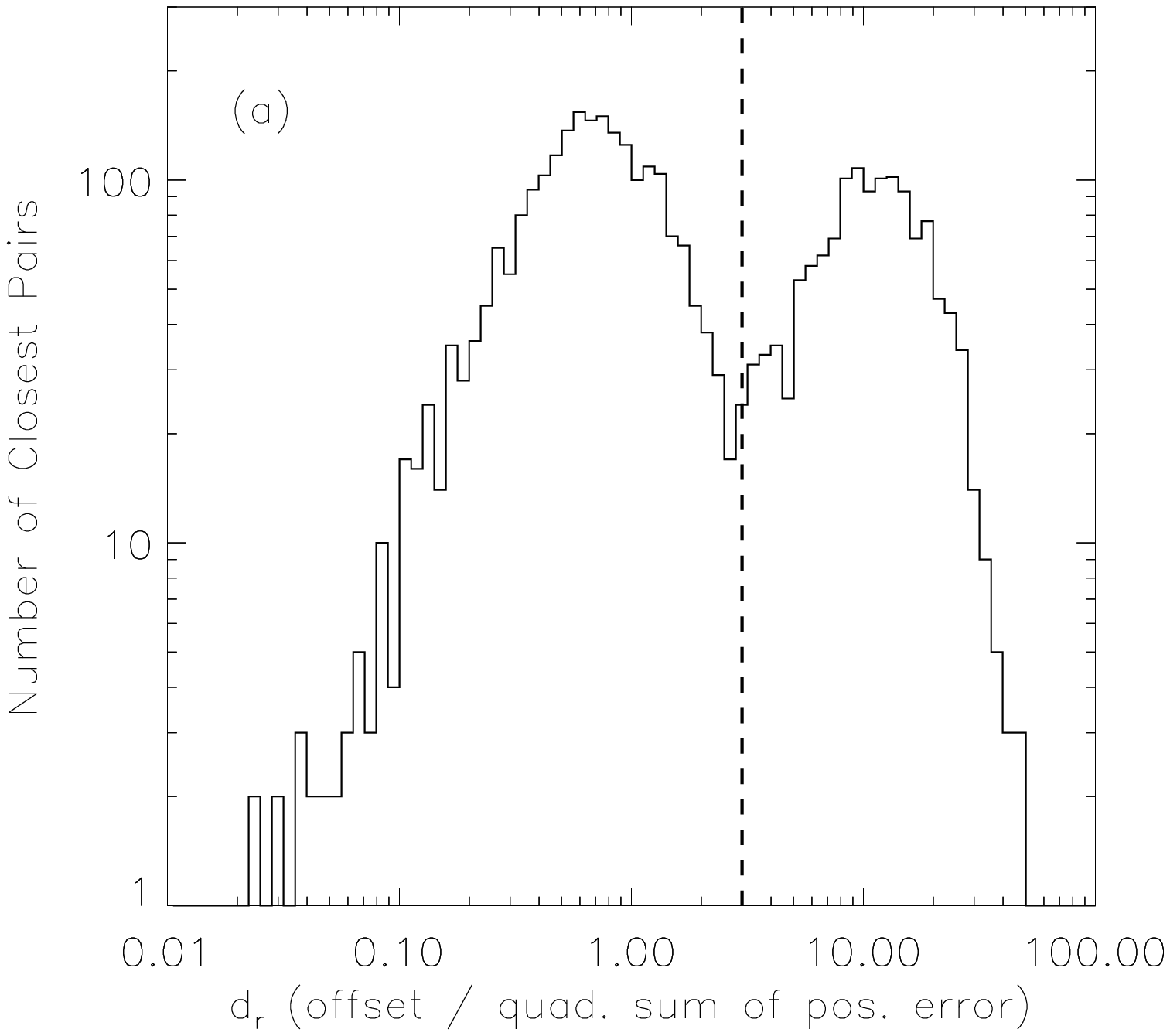}
       \includegraphics[scale=0.5]{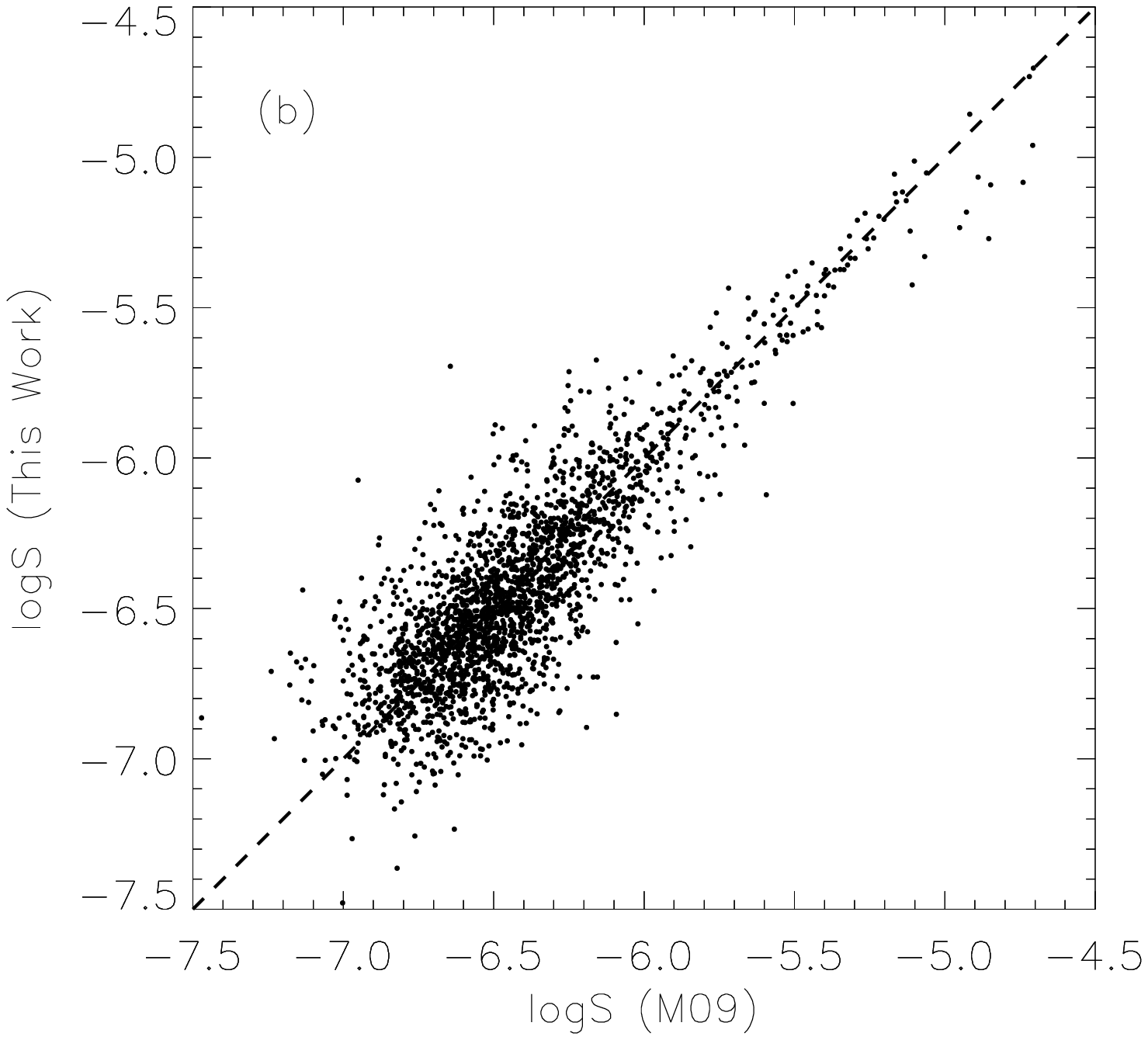}
	\caption{(a) The distribution of the relative distance between a M09 source and its closest pair in our main catalog. Pairs with a relative distance $<$ 3 (vertical dashed line) are considered true counterparts in the two catalogs. 
(b) Comparison of the photometry between M09 and this work for the common sources. The diagonal dashed line shows a 1:1 correspondence.}
	\label{fig:cc}
\end{figure}

\begin{figure}
	\centering
	\includegraphics[scale=0.8]{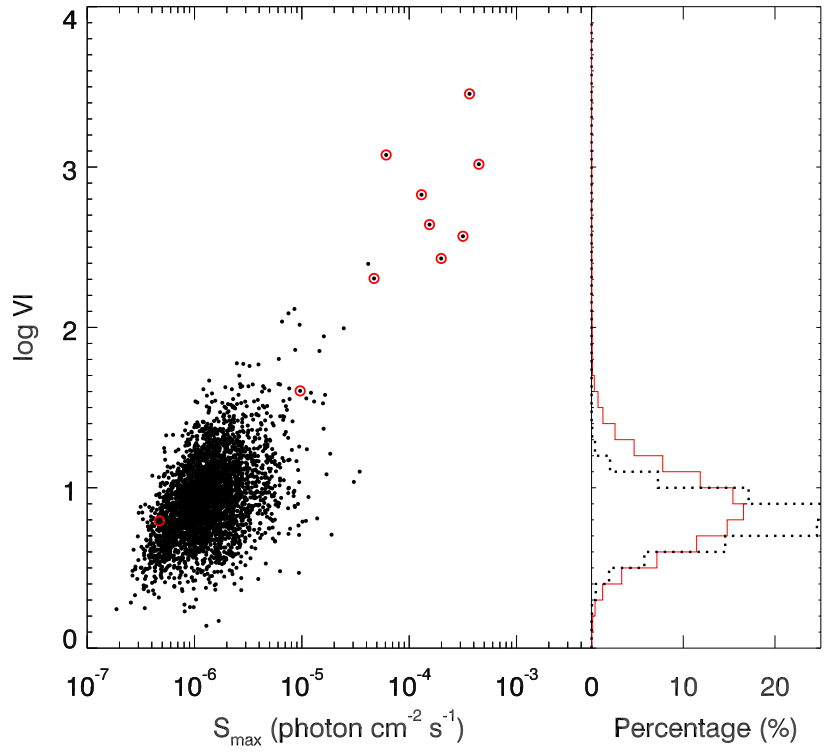}
	\caption{The X-ray variability index (VI) versus the maximum 2--8 photon flux. The red circles highlight previously known transients (\citealp{Degenaar2015}). The right panel shows the histogram of VI (solid red), 
in comparison with the simulated VI distribution of constant sources following pure Poisson fluctuations and a power-law luminosity function (dashed black). 
}
	\label{fig:var}
\end{figure}

\begin{figure}
	\centering
	\includegraphics[scale=0.45,angle=270]{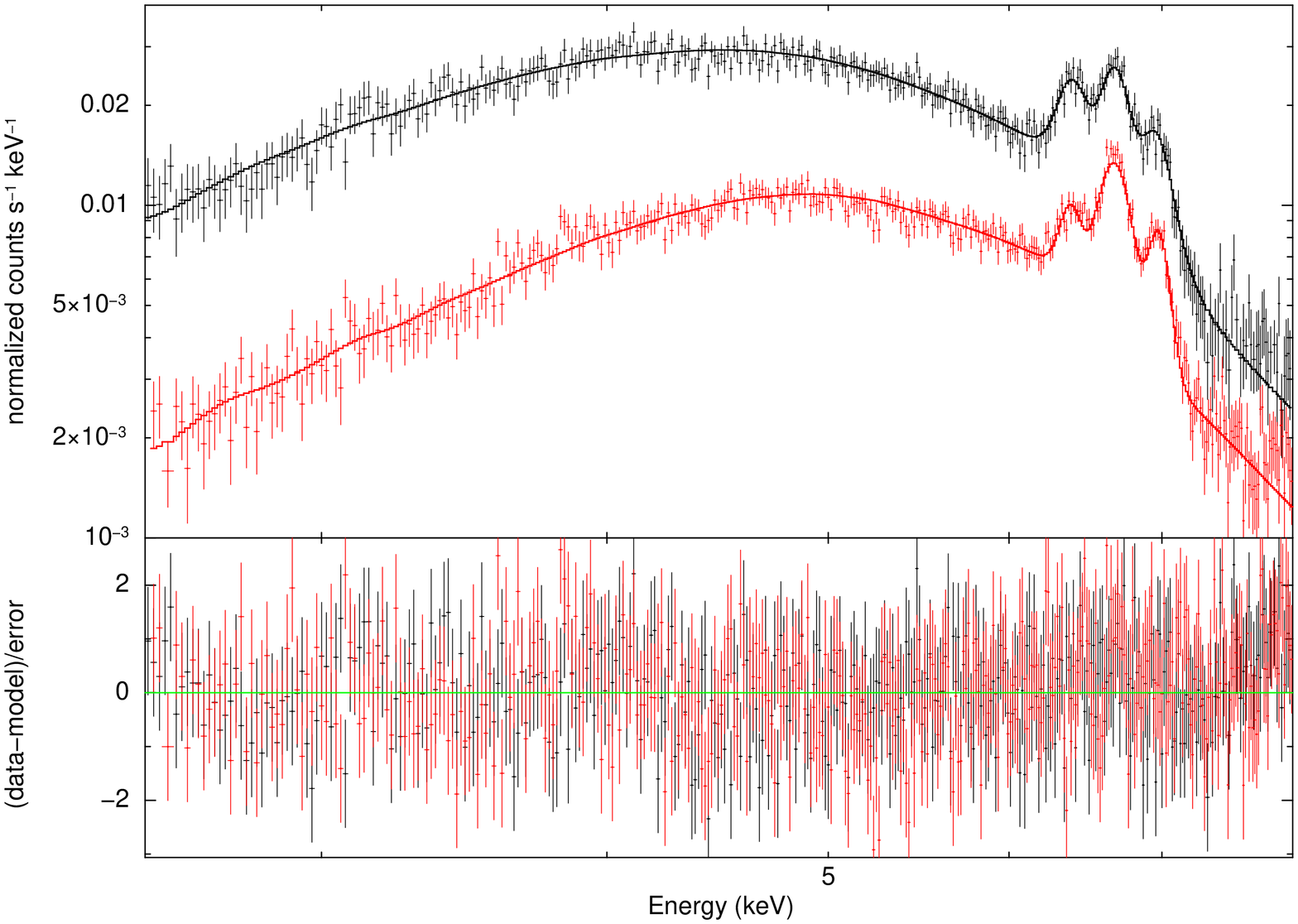}
	\\[15pt]
        \includegraphics[scale=0.45,angle=270]{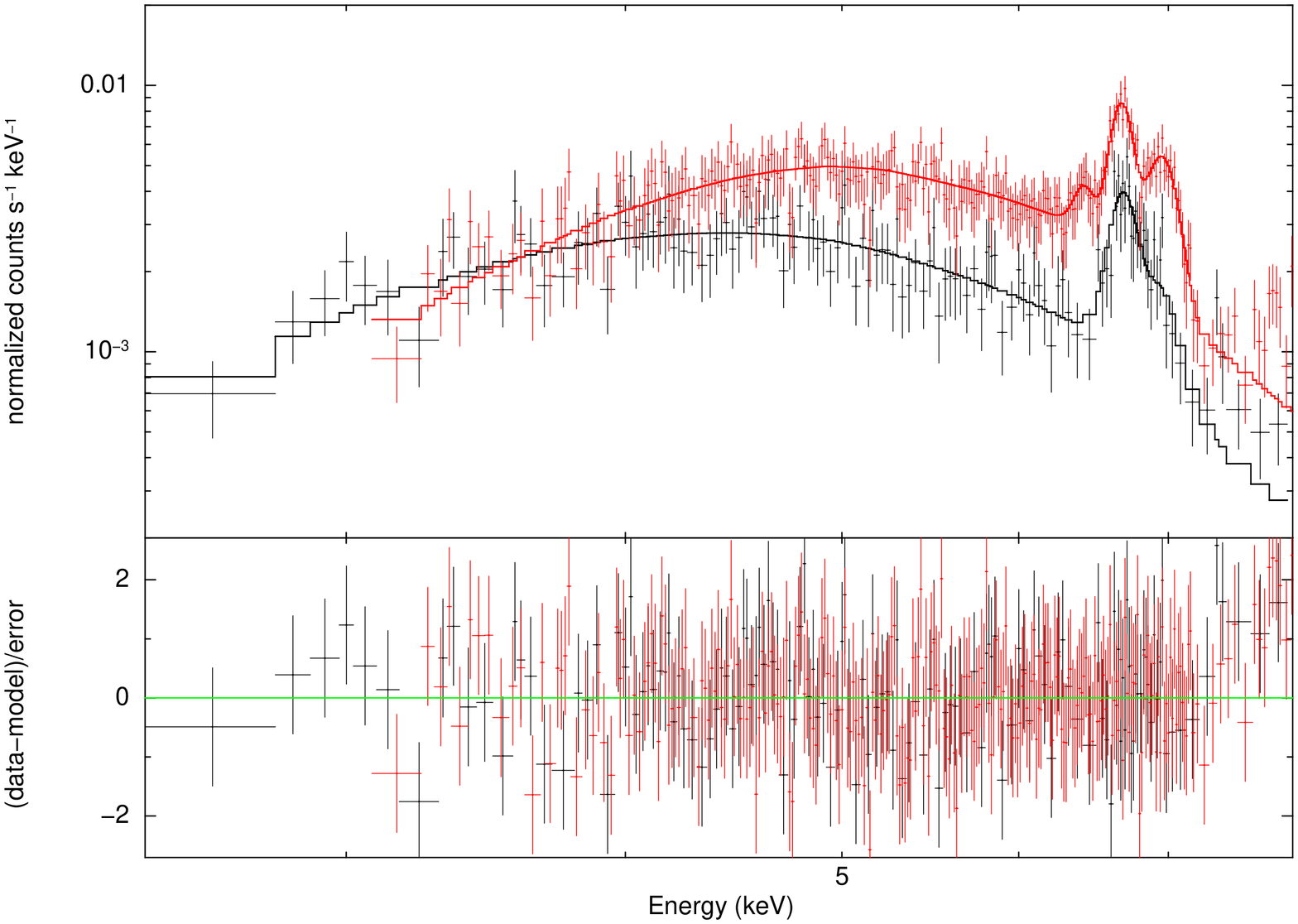}
	\caption{Cumulative source spectra with the best-fit model, an absorbed bremsstrahlung plus three Gaussians for the 6.4, 6.7 and 7.0 keV Fe lines. 
The ACIS-I and ACIS-S spectra are shown in black and red, respectively. 
The spectra have been adaptively binned to achieve S/N greater than 3.
The {\it upper panel} displays bright sources ($L_{\rm X} \gtrsim 6\times 10^{31}$~erg~s$^{-1}$), while the {\it lower panel} is for faint sources ($L_{\rm X} \lesssim 6\times 10^{31}$~erg~s$^{-1}$). }
	\label{fig:spec}
\end{figure}

\begin{figure}
	\centering
        \includegraphics[scale=0.5,angle=270]{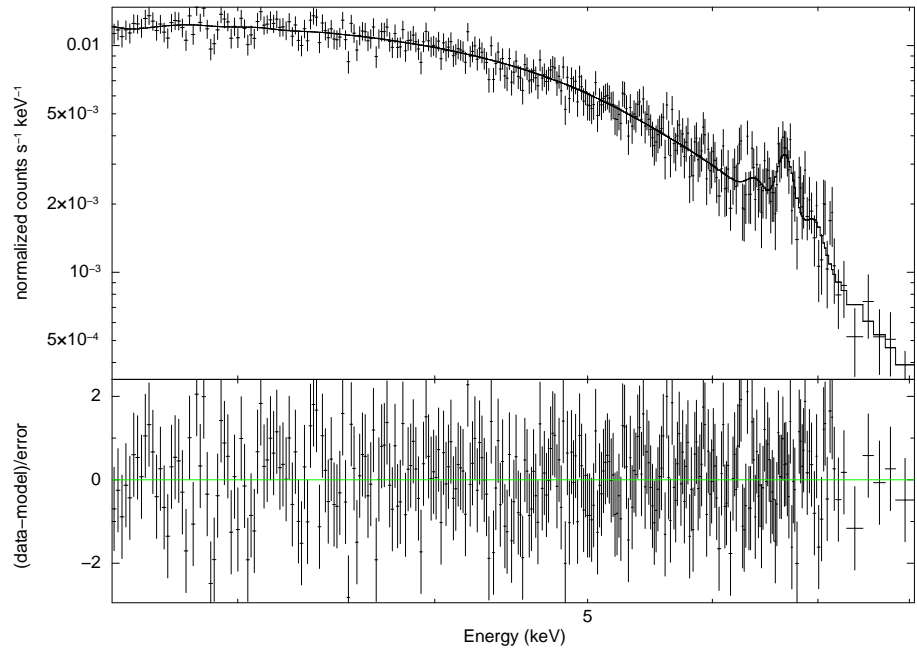}
	\caption{The cumulative spectrum of LW sources with the best-fit model as described in Figure \ref{fig:spec}. The spectrum has been adaptively binned to achieve S/N greater than 3. 
}
	\label{fig:spec_lw}
\end{figure}

\begin{figure}
      \centering
       \includegraphics[scale=0.55]{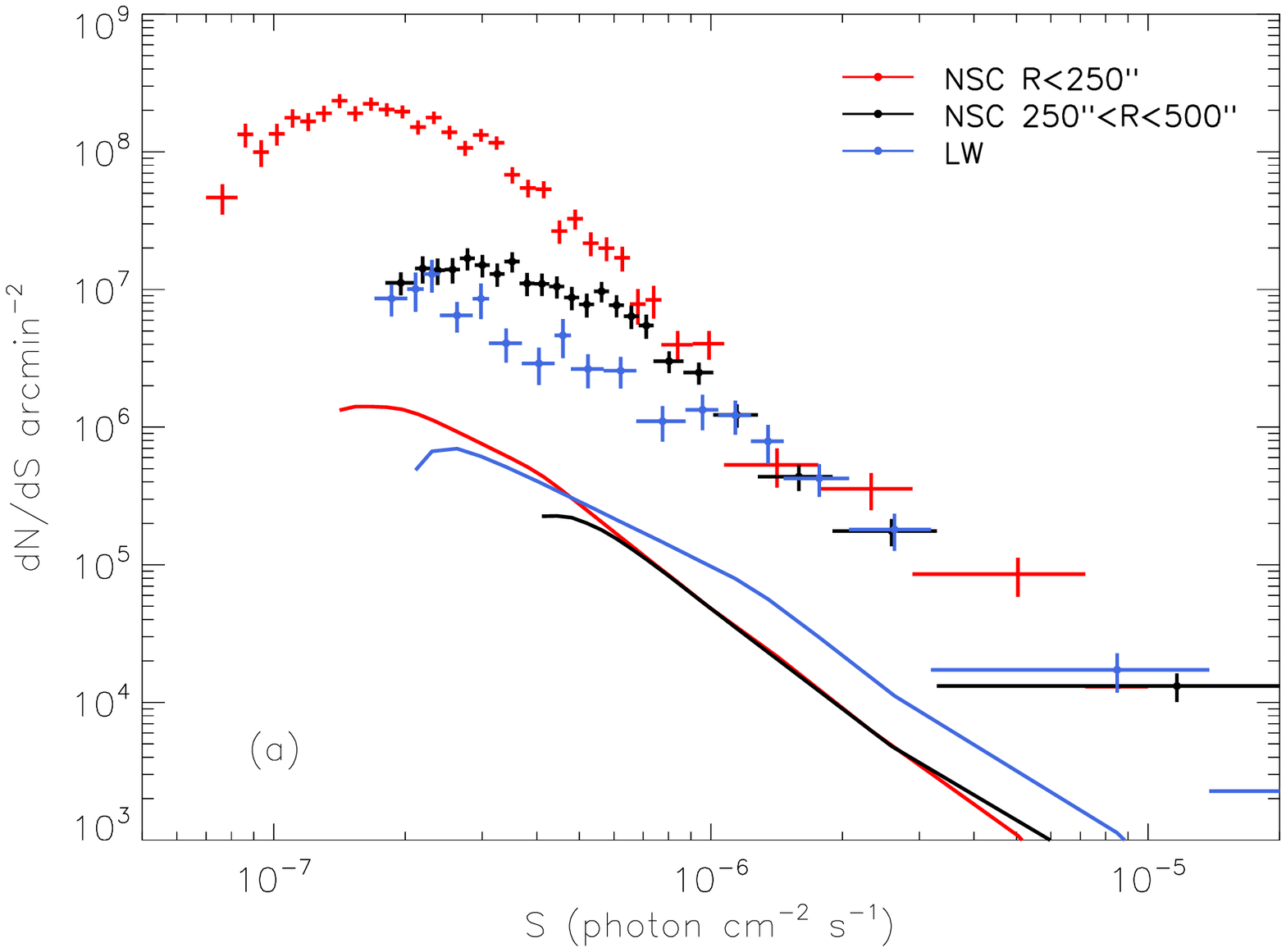}
       \\[10pt]
       \includegraphics[scale=0.55]{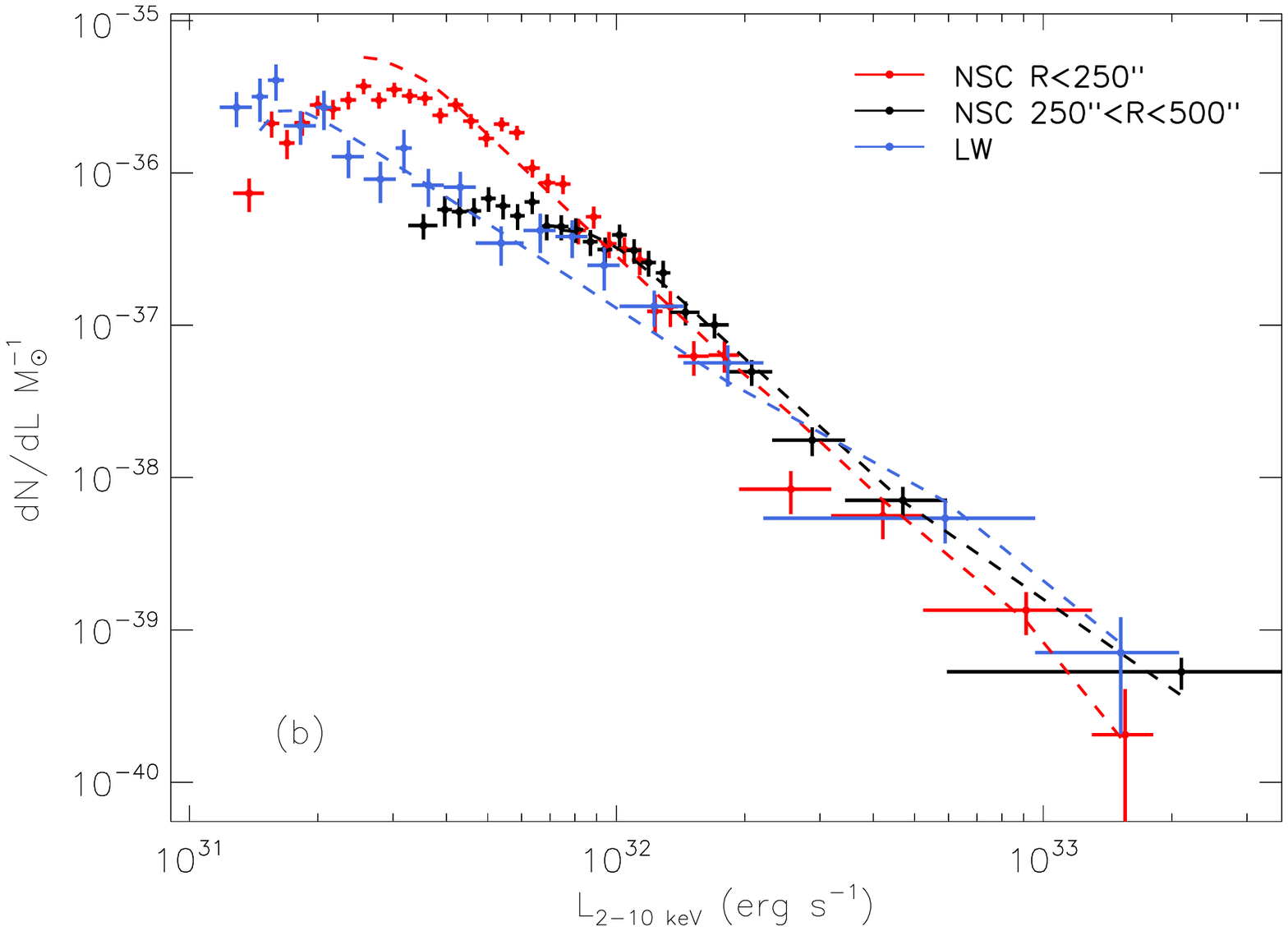}
	\caption{
{\it Upper panel}: Differential flux distributions of sources in the GC and LW,
normalized by the underlying sky area. Also plotted in solid curves are the $logN-logS$ relation of the CXB, which has a negligible contribution in all three cases.
{\it Lower panel}: Luminosity functions of the GC and LW sources, normalized by the enclosed stellar mass. 
The unabsorbed 2--10 keV luminosity is derived using the photon flux-to-energy flux conversion factors from the spectral analysis (Section~\ref{sec:spec}). The LW luminosity function (blue data points) is thus shifted to the left, due to a lower conversion factor. 
Also plotted in dashed curves are the best-fit models, including a predominant power-law for the GC/LW sources and a minor contribution from the CXB, both corrected for detection incompleteness and Eddington bias. The model is fitted to fluxes above the $\sim$50\% completeness in each case, to minimize systematics at lower fluxes .
}
	\label{fig:LF}
\end{figure}

\begin{figure}
	\centering
	\includegraphics[scale=0.7]{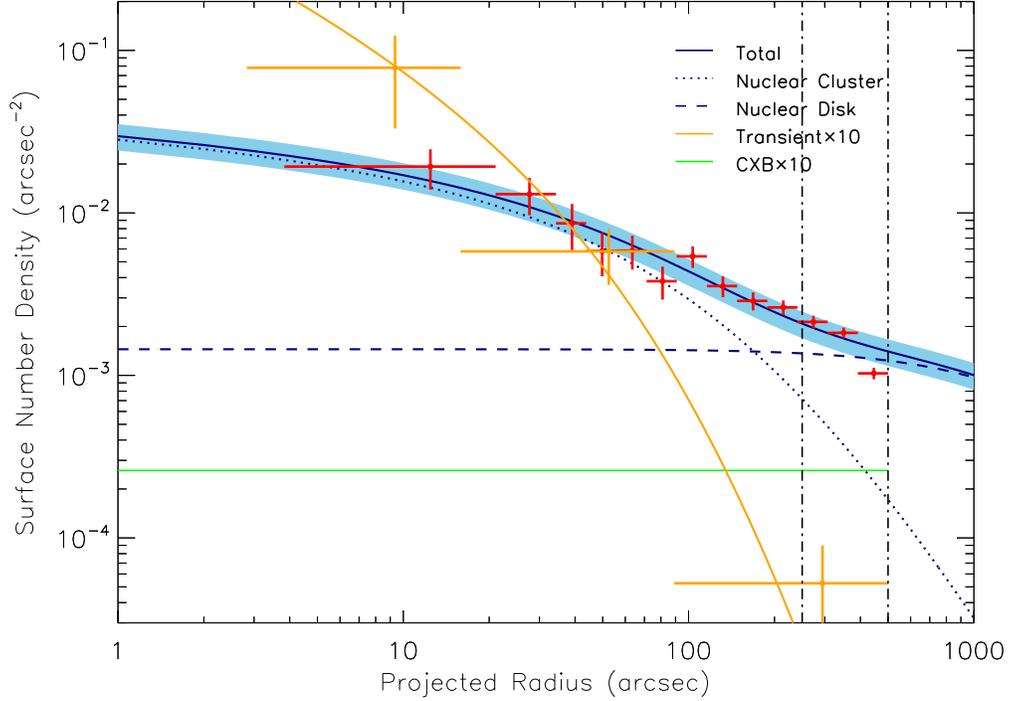}
	\caption{
Surface density profile of X-ray sources with $S_{2-8} >$ 3.5$\times10^{-7}$~photon~cm$^{-2}$~s$^{-1}$ (red points).
The solid curve is a model fitted to the surface density profile within $R < 250\arcsec$, with the shaded area representing the uncertainty range. The model, derived from the projected stellar mass distribution of \citet{Fritz2016}, consists of two components, the nuclear cluster (dotted curve) and the nuclear disk (dashed curve).
The surface density profile of known transients (orange points, multiplied by a factor of 10 for clarity) appears much steeper and can be approximately matched by a projected $\rho_s^2$ distribution (orange curve), where $\rho_s(r)$ is the radial stellar density distribution of the NSC.  
The negligible contribution of the CXB, multiplied by a factor of 10, is also plotted in green for illustration.
}	
\label{fig:spatial}
\end{figure}

\begin{figure}
        \centering
        \includegraphics[scale=0.5,angle=90]{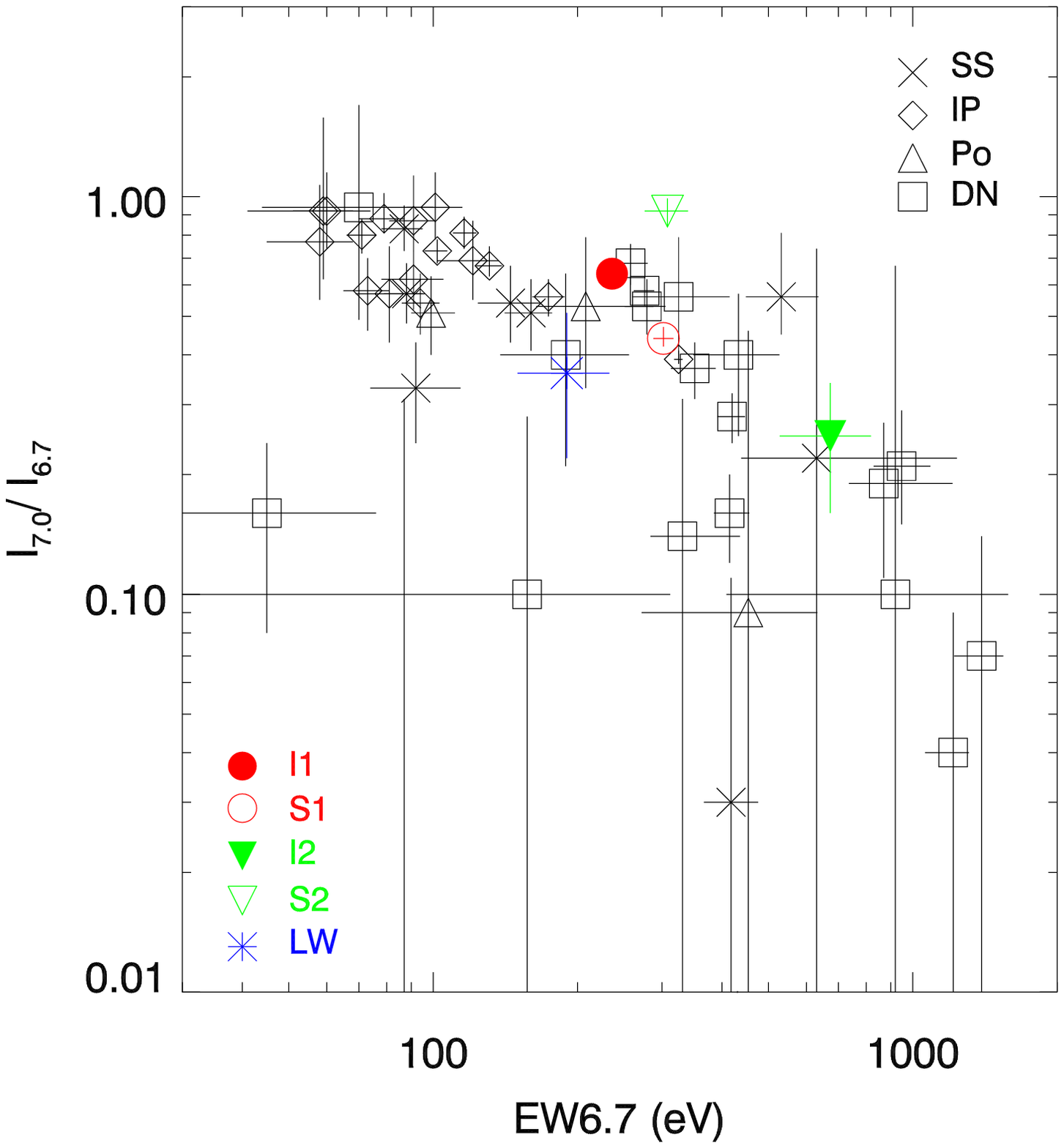}
        \includegraphics[scale=0.5,angle=90]{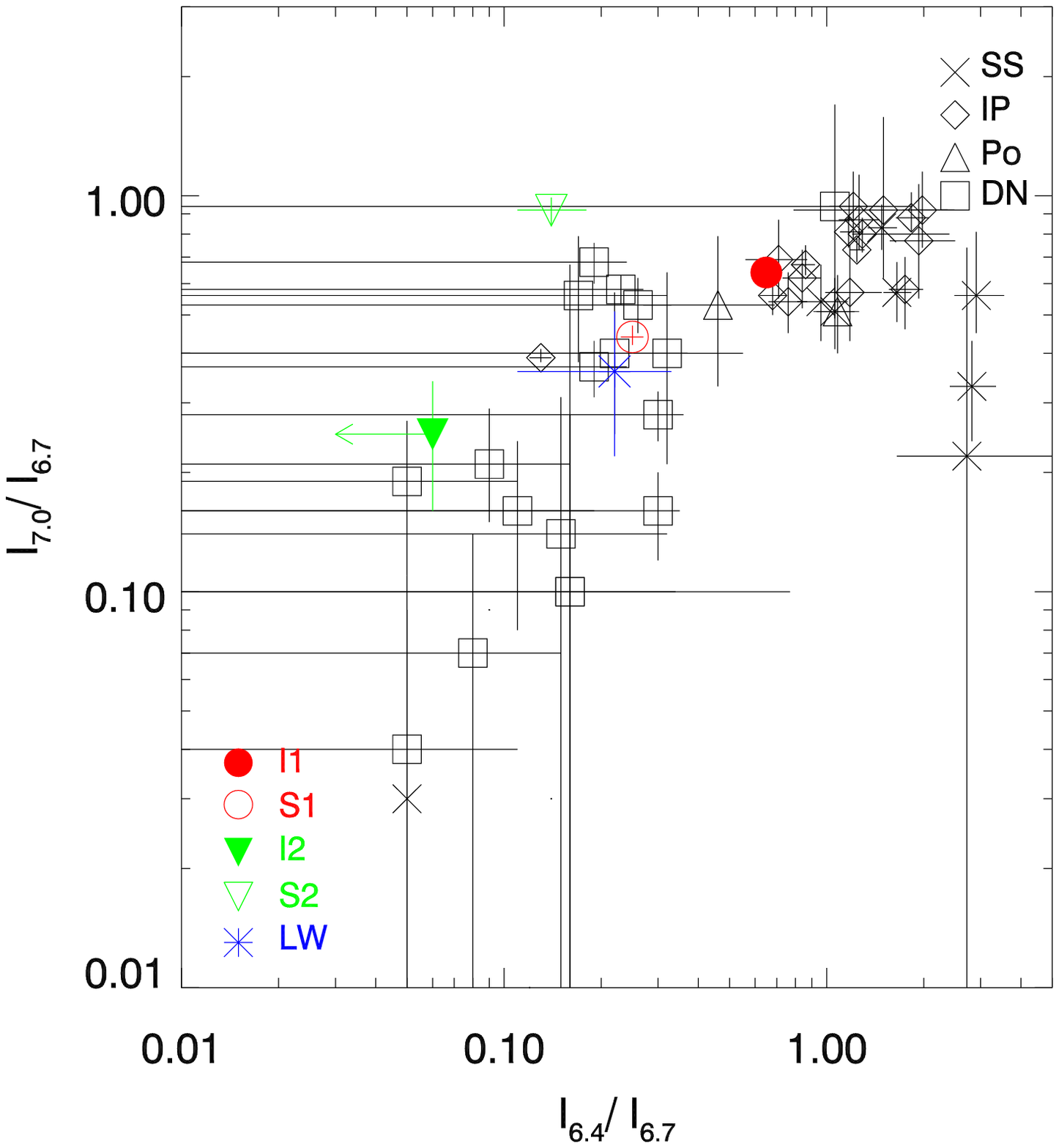}
        \caption{Flux ratio between the 7.0 and 6.7 keV lines versus (a) equivalent width of the 6.7 keV line, and (b) flux ratio between the 6.4 and 6.7 keV lines. The color-coded symbols are for the GC and LW spectra (see text for definition of the different sets). Black symbols are measurements in individual CVs in the Solar neighborhood (\citealp{Xu2016}). SS: symbiotic stars; IP: intermediate polars; Po: polars; DN: dwarf novae.}
\label{fig:lineratio}
\end{figure}


\bibliographystyle{apj}
\bibliography{a}
\newpage
\begin{deluxetable}{lcccccc}
\tabletypesize{\footnotesize}
\tablecaption{{\it Chandra} Observation Log\label{tab:obs}}
\tablehead{
\colhead{ObsID }&
\colhead{Start Time }&
\colhead{Exposure}&
\colhead{Instrument}  &
\multicolumn{2}{c}{Aim Point}&
\colhead{Roll Angle} \\\cline{5-6}
 & &  & & \colhead{R.A. }& \colhead{Dec.}&  \\
  & \colhead{(UT)} &\colhead{(ks)} & & \multicolumn{2}{c}{(J2000)}& \colhead{ (degree) }
}
\startdata
Galactic Center&&&&&&\\
242&1999-09-21  02:43:00& 45.9&ACIS-I&       266.41399& -29.01271&     268.7\\
1561a&2000-10-26  19:08:03& 35.8&ACIS-I&       266.41403& -29.01206&     264.7\\
1561b&2001-07-14  01:51:10& 13.5&ACIS-I&       266.41549& -29.01238&     280.7\\
2951&2002-02-19  14:27:32& 12.4&ACIS-I&       266.41862& -29.00345&      91.5\\
2952&2002-03-23  12:25:04& 11.9&ACIS-I&       266.41891& -29.00353&      88.2\\
2953&2002-04-19  10:59:43& 11.6&ACIS-I&       266.41916& -29.00364&      85.2\\
2954&2002-05-07  09:25:07& 12.5&ACIS-I&       266.41938& -29.00374&      82.1\\
2943&2002-05-22  23:19:42& 37.7&ACIS-I&       266.41991& -29.00406&      75.5\\
3663&2002-05-24  11:50:13& 38.0&ACIS-I&       266.41993& -29.00407&      75.5\\
3392&2002-05-25  15:16:03&166.7&ACIS-I&       266.41992& -29.00408&      75.5\\
3393&2002-05-28  05:34:44&158.0&ACIS-I&       266.41992& -29.00407&      75.5\\
3665&2002-06-03  01:24:37& 89.9&ACIS-I&       266.41992& -29.00407&      75.5\\
3549&2003-06-19  18:28:55& 24.8&ACIS-I&       266.42095& -29.01052&     346.8\\
4683&2004-07-05  22:33:11& 49.5&ACIS-I&       266.41606& -29.01240&     286.2\\
4684&2004-07-06  22:29:57& 49.5&ACIS-I&       266.41597& -29.01239&     285.4\\
5360&2004-08-28  12:03:59&  5.1&ACIS-I&       266.41477& -29.01214&     271.0\\
6113&2005-02-27  06:26:04&  4.9&ACIS-I&       266.41870& -29.00350&      90.6\\
5950&2005-07-24  19:58:27& 48.5&ACIS-I&       266.41519& -29.01225&     276.7\\
5951&2005-07-27  19:08:16& 44.6&ACIS-I&       266.41512& -29.01222&     276.0\\
5952&2005-07-29  19:51:11& 45.3&ACIS-I&       266.41508& -29.01222&     275.5\\
5953&2005-07-30  19:38:31& 45.4&ACIS-I&       266.41506& -29.01221&     275.3\\
5954&2005-08-01  20:16:05& 17.9&ACIS-I&       266.41503& -29.01218&     274.9\\
6639&2006-04-11  05:33:20&  4.5&ACIS-I&       266.41891& -29.00366&      86.2\\
6640&2006-05-03  22:26:26&  5.1&ACIS-I&       266.41935& -29.00380&      82.8\\
6641&2006-06-01  16:07:52&  5.1&ACIS-I&       266.42019& -29.00437&      69.7\\
6642&2006-07-04  11:01:35&  5.1&ACIS-I&       266.41634& -29.01240&     288.4\\
6363&2006-07-17  03:58:28& 29.8&ACIS-I&       266.41542& -29.01231&     279.5\\
6643&2006-07-30  14:30:26&  5.0&ACIS-I&       266.41510& -29.01221&     275.4\\
6644&2006-08-22  05:54:34&  5.0&ACIS-I&       266.41485& -29.01205&     271.7\\
6645&2006-09-25  13:50:35&  5.1&ACIS-I&       266.41448& -29.01197&     268.3\\
6646&2006-10-29  03:28:20&  5.1&ACIS-I&       266.41425& -29.01181&     264.4\\
7554&2007-02-11  06:16:55&  5.1&ACIS-I&       266.41846& -29.00332&      92.6\\
7555&2007-03-25  22:56:07&  5.1&ACIS-I&       266.41414& -29.00002&      88.0\\
7556&2007-05-17  01:05:03&  5.0&ACIS-I&       266.41556& -28.99973&      79.5\\
7557&2007-07-20  02:27:01&  5.0&ACIS-I&       266.42069& -29.01498&     278.4\\
7558&2007-09-02  20:19:41&  5.0&ACIS-I&       266.41945& -29.01543&     270.5\\
7559&2007-10-26  10:04:04&  5.0&ACIS-I&       266.41868& -29.01564&     264.8\\
9169&2008-05-05  03:53:16& 27.6&ACIS-I&       266.41522& -28.99981&      81.7\\
9170&2008-05-06  03:00:30& 26.8&ACIS-I&       266.41521& -28.99981&      81.7\\
9171&2008-05-10  03:18:02& 27.7&ACIS-I&       266.41522& -28.99980&      81.7\\
9172&2008-05-11  03:36:46& 27.4&ACIS-I&       266.41521& -28.99981&      81.7\\
9174&2008-07-25  21:50:50& 28.8&ACIS-I&       266.42039& -29.01521&     276.4\\
9173&2008-07-26  21:20:49& 27.8&ACIS-I&       266.42035& -29.01521&     276.2\\
10556&2009-05-18  02:19:58&112.5&ACIS-I&       266.41566& -28.99975&      79.0\\
11843&2010-05-13  02:12:34& 78.9&ACIS-I&       266.41539& -28.99977&      80.7\\
13016&2011-03-29  10:30:09& 17.8&ACIS-I&       266.41431& -28.99996&      87.6\\
13017&2011-03-31  10:30:09& 17.8&ACIS-I&       266.41435& -28.99998&      87.4\\
14941&2013-04-06  01:23:27& 19.8&ACIS-I&       266.41443& -28.99996&      86.8\\
14942&2013-04-14  15:42:54& 19.8&ACIS-I&       266.41459& -28.99996&      85.8\\
\hline
13850&2012-02-06  00:38:33& 59.3&ACIS-S&       266.41369& -29.00629&      92.2\\
14392&2012-02-09  06:18:08& 58.5&ACIS-S&       266.41369& -29.00628&      92.2\\
14394&2012-02-10  03:17:24& 17.8&ACIS-S&       266.41367& -29.00628&      92.2\\
14393&2012-02-11  10:14:08& 41.0&ACIS-S&       266.41369& -29.00629&      92.2\\
13856&2012-03-15  08:46:28& 39.5&ACIS-S&       266.41368& -29.00630&      92.2\\
13857&2012-03-17  08:58:50& 39.0&ACIS-S&       266.41369& -29.00628&      92.2\\
13854&2012-03-20  10:13:19& 22.8&ACIS-S&       266.41368& -29.00629&      92.2\\
14413&2012-03-21  06:45:56& 14.5&ACIS-S&       266.41367& -29.00630&      92.2\\
13855&2012-03-22  11:25:56& 19.8&ACIS-S&       266.41369& -29.00628&      92.2\\
14414&2012-03-23  17:49:44& 19.8&ACIS-S&       266.41366& -29.00629&      92.2\\
13847&2012-04-30  16:17:58&152.0&ACIS-S&       266.41426& -29.00563&      76.6\\
14427&2012-05-06  20:02:07& 79.0&ACIS-S&       266.41426& -29.00563&      76.4\\
13848&2012-05-09  12:03:55& 96.9&ACIS-S&       266.41427& -29.00562&      76.4\\
13849&2012-05-11  03:19:47&176.4&ACIS-S&       266.41427& -29.00563&      76.4\\
13846&2012-05-16  10:42:22& 55.5&ACIS-S&       266.41426& -29.00562&      76.4\\
14438&2012-05-18  04:29:45& 25.5&ACIS-S&       266.41427& -29.00561&      76.4\\
13845&2012-05-19  10:43:37&133.5&ACIS-S&       266.41427& -29.00563&      76.4\\
14460&2012-07-09  22:34:10& 23.7&ACIS-S&       266.41991& -29.00884&     282.3\\
13844&2012-07-10  23:12:04& 19.8&ACIS-S&       266.41991& -29.00884&     282.3\\
14461&2012-07-12  05:49:52& 50.3&ACIS-S&       266.41991& -29.00885&     282.3\\
13853&2012-07-14  00:38:24& 72.7&ACIS-S&       266.41991& -29.00885&     282.3\\
13841&2012-07-17  21:07:45& 44.5&ACIS-S&       266.41992& -29.00885&     282.3\\
14465&2012-07-18  23:24:45& 43.8&ACIS-S&       266.41992& -29.00886&     282.3\\
14466&2012-07-20  12:38:16& 44.5&ACIS-S&       266.41991& -29.00884&     282.3\\
13842&2012-07-21  11:53:47&189.3&ACIS-S&       266.41991& -29.00885&     282.3\\
13839&2012-07-24  07:04:06&173.9&ACIS-S&       266.41991& -29.00885&     282.3\\
13840&2012-07-26  20:02:58&160.4&ACIS-S&       266.41991& -29.00885&     282.3\\
14432&2012-07-30  12:57:08& 73.3&ACIS-S&       266.41992& -29.00885&     282.3\\
13838&2012-08-01  17:30:32& 98.3&ACIS-S&       266.41991& -29.00885&     282.3\\
13852&2012-08-04  02:38:43&154.5&ACIS-S&       266.41991& -29.00885&     282.3\\
14439&2012-08-06  22:18:06&110.3&ACIS-S&       266.41960& -29.00940&     270.7\\
14462&2012-10-06  16:33:00&131.6&ACIS-S&       266.41953& -29.00949&     268.7\\
14463&2012-10-16  00:53:35& 30.4&ACIS-S&       266.41954& -29.00948&     268.7\\
13851&2012-10-16  18:49:52&105.7&ACIS-S&       266.41953& -29.00949&     268.7\\
15568&2012-10-18  08:56:30& 35.6&ACIS-S&       266.41954& -29.00950&     268.7\\
13843&2012-10-22  16:01:55&119.1&ACIS-S&       266.41954& -29.00949&     268.7\\
15570&2012-10-25  03:31:50& 67.8&ACIS-S&       266.41953& -29.00949&     268.7\\
14468&2012-10-29  23:43:14&144.2&ACIS-S&       266.41954& -29.00949&     268.7\\
\hline\hline
Limiting Window&&&&&&\\
5934&2005-08-22  08:17:36& 40.5&ACIS-I&       267.86692& -29.59236&     272.7\\
6362&2005-08-19  15:48:02& 37.7&ACIS-I&       267.86694& -29.59238&     273.1\\
6365&2005-10-25  14:37:40& 20.7&ACIS-I&       267.86629& -29.59215&     265.3\\
9500&2008-07-20  07:43:13&162.6&ACIS-I&       267.86148& -29.58793&     280.0\\
9501&2008-07-23  07:56:39&131.0&ACIS-I&       267.86399& -29.58953&     278.9\\
9502&2008-07-17  15:22:25&164.1&ACIS-I&       267.86685& -29.59108&     281.2\\
9503&2008-07-28  17:20:45&102.3&ACIS-I&       267.86852& -29.59311&     275.2\\
9504&2008-08-02  21:08:11&125.4&ACIS-I&       267.87097& -29.59490&     275.2\\
9505&2008-05-07  15:07:59& 10.7&ACIS-I&       267.85740& -29.57123&      82.2\\
9854&2008-07-27  05:30:09& 22.8&ACIS-I&       267.87404& -29.59630&     277.7\\
9855&2008-05-08  04:42:41& 55.9&ACIS-I&       267.85741& -29.57124&      82.2\\
9892&2008-07-31  07:50:48& 65.8&ACIS-I&       267.86853& -29.59312&     275.2\\
9893&2008-08-01  02:39:25& 42.2&ACIS-I&       267.87098& -29.59490&     275.2\\

\enddata
\end{deluxetable}

%

\begin{deluxetable}{l  c c c c c c c c r} 
\tabletypesize{\footnotesize}
\tablecaption{Catalog of the $R < 500\arcsec$ sources detected in the 0.5-2 keV band\label{tab:soft}}
\tablehead{
\colhead{No}&
\colhead{R.A.}&
\colhead{Dec.}&
\colhead{Pos. err}&
\colhead{$C_{t,0.5-2}$}&
\colhead{$C_{b,0.5-2}$}&
\colhead{$C_{net,0.5-2}$}&
\colhead{$S_{0.5-2}$}&
\colhead{HR0}\\
\colhead{(1)}&
\colhead{(2)}&
\colhead{(3)}&
\colhead{(4)}&
\colhead{(5)}&
\colhead{(6)}&
\colhead{(7)}&
\colhead{(8)}&
\colhead{(9)}
}
\startdata
       1&       266.26885& -29.12908&     0.203334&        3016&     167.9&
$2851.2^{+51.5}_{-57.3}$&     61.84&$-0.824^{+0.013}_{-0.013}$\\
       2&       266.27147& -29.11787&     0.127196&       12560&     162.1&
$12390.9^{+107.7}_{-112.2}$&    262.80&$-0.970^{+0.003}_{-0.003}$\\
       3&       266.27436& -29.10707&     0.254185&        1043&     126.8&
$923.9^{+22.2}_{-42.5}$&     24.18&$-0.377^{+0.028}_{-0.032}$\\
       4&       266.27475& -29.09783&     0.154812&         977&      84.2&
$889.4^{+34.0}_{-27.3}$&     16.82&$-0.256^{+0.028}_{-0.027}$\\
       5&       266.27732& -29.08953&     0.167426&         436&      67.8&
$367.0^{+21.9}_{-19.0}$&      7.17&$-0.578^{+0.049}_{-0.058}$\\
       6&       266.27901& -29.08238&    0.0886353&        4230&      42.5&
$4182.7^{+72.8}_{-55.2}$&     81.79&$-0.949^{+0.006}_{-0.006}$\\
       7&       266.28181& -29.08108&     0.172090&         823&      73.4&
$752.8^{+25.3}_{-31.6}$&     17.29&$-0.811^{+0.030}_{-0.034}$\\
       8&       266.28459& -29.07893&     0.144365&         762&      62.8&
$695.1^{+30.5}_{-23.3}$&     13.66&$-0.886^{+0.029}_{-0.034}$\\
       9&       266.28515& -29.07797&     0.158150&         296&      43.9&
$248.3^{+20.9}_{-13.2}$&      4.71&$-0.712^{+0.052}_{-0.069}$\\
      10&       266.28571& -29.06626&     0.106089&        1545&      44.3&
$1502.2^{+41.9}_{-35.6}$&     28.46&$-0.699^{+0.019}_{-0.019}$\\

\enddata
\tablecomments{(1) Source sequence number assigned in order of increasing R.A.; (2)-(3) Right Ascension and Declination (J2000) of source centroid; (4) Positional uncertainty, in arcseconds; (5)-(7) The total, background and net counts in the 0.5--2 keV band; 
(8) The 0.5--2 keV photon flux, in units of $10^{-7}$ photons~cm$^{-2}$~s$^{-1}$; (9) The soft color and 1\,$\sigma$ uncertainties. 
(Only a portion of the full table is shown here to illustrate its form and content.)
}
\end{deluxetable}
\clearpage
\begin{landscape}
\begin{deluxetable}{l c c c c c c c c c c c c c c c c r} 
\tabletypesize{\scriptsize}
\tablewidth{-20pt}
\tablecaption{Catalog of the $R < 500\arcsec$ sources detected in the 2-8 keV band\label{tab:main}}
\tablehead{
\colhead{No}&
\colhead{R.A.}&
\colhead{Dec.}&
\colhead{Pos. err}&
\colhead{$C_{t,2-8}$}&
\colhead{$C_{b,2-8}$}&
\colhead{$C_{net,2-8}$}&
\colhead{$S_{0.5-2}$}&
\colhead{$S_{2-8}$}&
\colhead{$F_{2-8}$}&
\colhead{HR0}&
\colhead{HR1}&
\colhead{HR2}&
\colhead{Note.}\\
\colhead{(1)}&
\colhead{(2)}&
\colhead{(3)}&
\colhead{(4)}&
\colhead{(5)}&
\colhead{(6)}&
\colhead{(7)}&
\colhead{(8)}&
\colhead{(9)}&
\colhead{(10)}&
\colhead{(11)}&
\colhead{(12)}&
\colhead{(13)}&
\colhead{(14)}&\\
}
\startdata
       1&       266.26001& -28.99571&      0.66&         152&      90.6&
$55.1^{+13.0}_{-11.9}$&      0.56&      6.12&     14.57&$>0.357$&
$0.306^{+0.241}_{-0.277}$&$-0.272^{+0.339}_{-0.247}$&...\\
       2&       266.26126& -29.02393&      0.38&         627&     387.9&
$263.4^{+30.8}_{-23.2}$&      1.76&     12.56&     29.90&
$0.127^{+0.189}_{-0.192}$&$0.161^{+0.157}_{-0.141}$&$0.045^{+0.142}_{-0.118}$&m
\\
       3&       266.26188& -28.99004&      0.60&         459&     371.9&
$84.7^{+26.9}_{-21.1}$&      0.20&      8.03&     19.12&
$0.166^{+0.834}_{-0.338}$&$>0.482$&$0.021^{+0.283}_{-0.286}$&m\\
       4&       266.26208& -29.01484&      0.39&         275&     116.5&
$165.1^{+21.3}_{-12.4}$&      1.33&      8.95&     21.31&
$0.255^{+0.196}_{-0.178}$&$0.230^{+0.128}_{-0.145}$&$0.005^{+0.122}_{-0.117}$&m
\\
       5&       266.26381& -28.99691&      0.60&         162&      92.9&
$62.7^{+16.9}_{-9.5}$&      0.46&      5.20&     12.38&$0.535^{+0.465}_{-0.134}$
&$-0.052^{+0.285}_{-0.305}$&$0.227^{+0.250}_{-0.221}$&...\\
       6&       266.26532& -28.96850&      0.40&         573&     389.2&
$179.8^{+30.6}_{-21.9}$&      0.46&     17.64&     41.98&$>0.327$&
$0.495^{+0.217}_{-0.324}$&$0.145^{+0.157}_{-0.128}$&m\\
       7&       266.26590& -29.00131&      0.36&         630&     406.0&
$245.5^{+29.5}_{-27.9}$&      1.62&     10.47&     24.92&
$-0.103^{+0.274}_{-0.220}$&$0.442^{+0.132}_{-0.199}$&$-0.175^{+0.150}_{-0.125}$&
m\\
       8&       266.26624& -29.01438&      0.24&        1530&     159.4&
$1379.7^{+46.7}_{-31.2}$&      0.96&     61.79&    147.07&
$0.828^{+0.066}_{-0.066}$&$0.498^{+0.034}_{-0.033}$&$-0.056^{+0.031}_{-0.029}$&m
\\
       9&       266.26661& -29.02758&      0.27&        1010&     189.9&
$808.2^{+34.6}_{-27.8}$&      1.23&     34.63&     82.43&
$0.736^{+0.083}_{-0.081}$&$0.334^{+0.049}_{-0.044}$&$-0.088^{+0.044}_{-0.043}$&m
\\
      10&       266.26907& -29.02754&      0.35&         379&     187.6&
$180.4^{+24.3}_{-15.9}$&      0.28&      8.42&     20.03&$>-0.298$&$>0.770$&
$0.069^{+0.092}_{-0.108}$&...\\

\enddata
\tablecomments{See Section \ref{sec:detail} for definition of the columns. 
(Only a portion of the full table is shown here to illustrate its form and content.)
}
\end{deluxetable}
\clearpage
\end{landscape}

\begin{deluxetable}{c p{1.6cm} p{1.3cm} p{1.3cm} p{1.3cm}  p{1.3cm} p{1.3cm} p{1.3cm} p{1.3cm} p{1cm} r} 
\centering
\tabletypesize{\footnotesize}
\tablewidth{-50pt}
\tablecaption{Spectral fit results}
\tablehead{
\colhead{Set}&
\colhead{$N_{\rm H_{wabs}}$}&
\colhead{$N_{\rm H_{pcf}}$}&
\colhead{$T_{\rm b}$}&
\colhead{CF}&
\colhead{EW$_{6.4}$}&
\colhead{EW$_{6.7}$}&
\colhead{EW$_{7.0}$}&
\colhead{$I_{6.4}/I_{6.7}$}&
\colhead{$I_{7.0}/I_{6.7}$}&
\colhead{$\chi^{2}/dof$}\\
&
\multicolumn{2}{c}{($10^{22}{\rm~cm^{-2}}$)}&
\colhead{(keV)}&
&
\colhead{(eV)}&
\colhead{(eV)}&
\colhead{(eV)}&
& &
}
\startdata
I1&$6.06^{+1.81}_{-2.94}$&$17.91^{+5.23}_{-2.79}$&$40 (fixed)$&$0.67^{+0.15}_{-0.12}$&$137\pm11$&$236\pm14$&$165^{+15}_{-16}$&$0.65_{-0.04}^{+0.05} $&$0.64^{+0.04}_{-0.05}$& 262.47/363\\

I2&$6.06 (fixed)$&$12.4^{+9.50}_{-6.70}$&$40 (fixed)$&$0.74^{+0.11}_{-0.11}$&$<40$&$673\pm145$&$108^{+60}_{-56}$&$<0.06$ &$ 0.25^{+0.09}_{-0.09}   $& 111.96/134\\

S1 &$7.85^{+1.41}_{-2.20}$&$20.74^{+6.77}_{4.70}$&$40 (fixed)$&$0.56^{+0.15}_{-0.10}$&$66^{+6}_{-7}$&$302^{+15}_{-14}$&$140\pm10$&$ 0.25^{+0.02}_{-0.02} $&$ 0.42^{+0.03}_{-0.04} $ & 367.04/367\\

S2 &$7.85 (fixed)$&$10.1^{+3.4}_{-1.0}$&$40 (fixed)$&$>0.85$&$41\pm17$&$308\pm32$&$310^{+37}_{-35}$& $0.14^{+0.04}_{-0.03}$&$ 0.92^{+0.07}_{-0.08}  $& 179.06/258\\

LW & $0.7 (fixed)$ & $0.62^{+0.1}_{-0.1}$&$17.2^{+11.2}_{-7.8}$&$>0.67$&$42^{+25}_{-26}$&$190^{+43}_{-40}$&$73^{+37}_{-38}$&$0.22^{+0.11}_{-0.10}$&$0.36^{+0.15}_{-0.14}$&$275.16/317$

\enddata
\label{tab:line}
\end{deluxetable}
\clearpage

\appendix
\section{Illustration of spurious and new sources}
In Figure~\ref{fig:panel1}, we show examples of sources identified by M09 but not included in our catalog. These primarily include extended features and background fluctuations (i.e., $P_B > 0.1$).  
In Figure \ref{fig:panel2}, we show examples of sources identified by this work but not included in M09. These primarily include transients and faint sources only becoming detectable due to the observations taken since the work of M09.     

For ease of reference, Table~\ref{tab:del_src} provides the approximate centroid positions of the extended sources and spurious sources (i.e, $P_B > 0.1$), which were originally resulted from {\it wavdetect}. 

\setcounter{figure}{0}
\renewcommand\thefigure{A\arabic{figure}}
\begin{figure}
	\centering
       \includegraphics[scale=0.9]{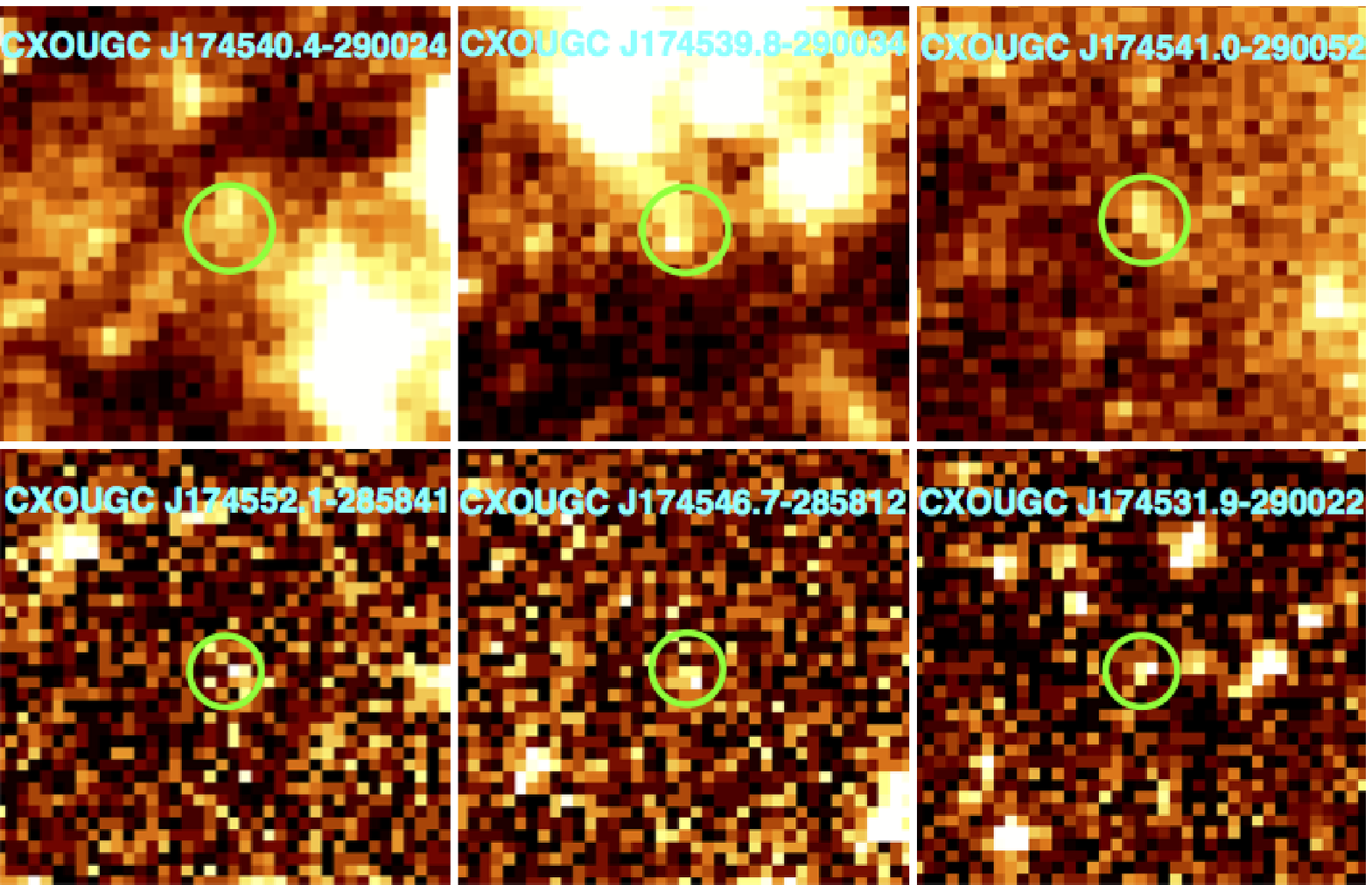}
	\caption{Zoom-in 2--8 keV counts image, with circles denoting the positions of M09 sources overlaid. The upper panel represents examples of M09 sources that are flagged as extended sources in our catalog. The lower panel shows examples of M09 sources that are identified as spurious sources from our catalog.}
	\label{fig:panel1}
\end{figure}

\begin{figure}
	\centering
       \includegraphics[scale=0.9]{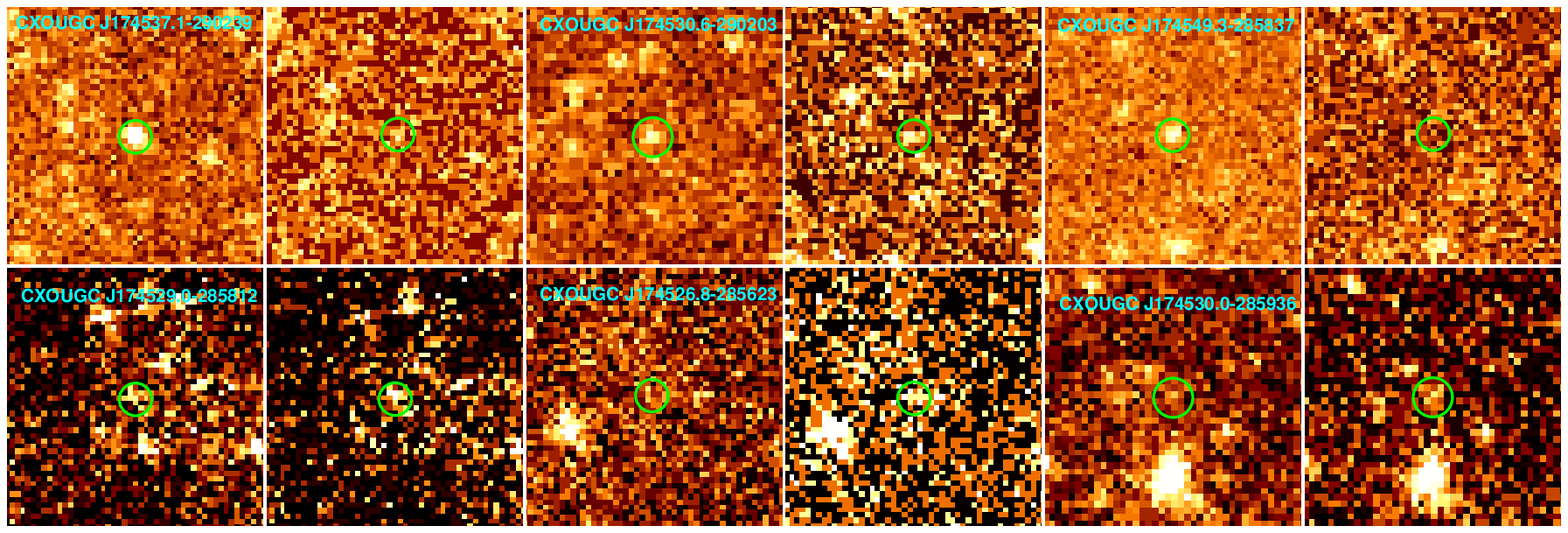}
	\caption{Examples of newly identified sources as seen in the data used this work and data in M09. The upper panel shows examples of recently brightened sources, while the lower panel shows examples of faint sources that were undetectable in M09. In both panels, each pair illustrates one source and its vicinity as appearing in our full dataset on the left, and the same region as in the dataset of M09 on the right.}
	\label{fig:panel2}
\end{figure}

\clearpage

\setcounter{table}{0}
\renewcommand\thetable{A\arabic{table}}

\begin{deluxetable}{lcccr}
\centering
\tabletypesize{\footnotesize}
\tablewidth{-5pt}
\tablecaption{Positions of removed sources originally found by {\it wavdetect}}
\tablehead{
\colhead{No}&
\colhead{R.A.}&
\colhead{Dec.}&
\colhead{Note.}\\
\colhead{(1)}&
\colhead{(2)}&
\colhead{(3)}&
\colhead{(4)}
}
\startdata
       1&       266.26906& -28.96497&S\\
       2&       266.27388& -28.96352&S\\
       3&       266.27715& -29.01983&S\\
       4&       266.28079& -29.07465&S\\
       5&       266.28350& -28.98203&S\\
       6&       266.29255& -29.01503&S\\
       7&       266.29903& -28.94985&S\\
       8&       266.30452& -29.02352&S\\
       9&       266.31060& -29.04073&S\\
      10&       266.31330& -28.92473&S\\
\enddata
\label{tab:del_src}
\tablecomments{(1) Source sequence number assigned in order of increasing R.A.; (2)-(3) Right Ascension and Declination (J2000) of source centroid;  (4) Classification of the removed sources, S for spurious sources and E for extended sources. (Only a portion of the full table is shown here to illustrate its form and content.)
}
\end{deluxetable}

\end{document}